  \documentclass[12pt]{article}
 \usepackage{epsfig}
\newcommand{\be}{\begin{equation}}
\newcommand{\ee}{\end{equation}}
\newcommand{\ba}{\begin{eqnarray}}
\newcommand{\ea}{\end{eqnarray}}

\newcommand{\bra}{\langle}
\newcommand{\ket}{\rangle}
\newcommand{\inc}{{\it i}}
\newcommand{\pbold}{\mbox{{\boldmath $
 p$}}}
\newcommand{\wbold}{\mbox{{\boldmath $
 w$}}}

\newcommand{\f}{{\mbox{\boldmath$
 f$}}}

\newcommand{\omegabold}{\mbox{{\boldmath $
{\omega}$}}}
\newcommand{\erbold}{\mbox{{\boldmath $
 {r}$}}}
\newcommand{\efbold}{\mbox{{\boldmath $
 {f}$}}}
\newcommand{\fbold}{\mbox{{\boldmath $
 {f}$}}}
\newcommand{\Fbold}{\mbox{{\boldmath $
 {F}$}}}

\newcommand{\Phibold}{\mbox{{\boldmath $
 {\Phi}$}}}
\newcommand{\mubold}{\mbox{{\boldmath $
 {\mu}$}}}

 \newcommand{\dotmubold}{\dot{\bf {\mbox{\boldmath $
 {\boldmath \mu}$}} }}

\newcommand{\doterbold}{\mbox{{\boldmath $\dot{
 {r}}$}}}
\newcommand{\ddoterbold}{\mbox{{\boldmath $\ddot{
 {r}}$}}}
\begin{document}

\title{
 Gauge Freedom in Orbital Mechanics.}
 \author{ {\Large{Michael Efroimsky}}\\
 {\small{US Naval Observatory, Washington DC 20392 USA}}\\
   }
 \maketitle
  \begin{abstract}
  {\small
 Both orbital and attitude dynamics employ the method of variation
of parameters. In a non-perturbed setting, the coordinates (or the
Euler angles) get expressed as functions of the time and six
adjustable constants called elements. Under disturbance, each such
expression becomes ansatz, the ``constants" being endowed with
time dependence. The perturbed velocity (linear or angular)
consists of a partial time derivative and a convective term
containing time derivatives of the ``constants." It can be shown
that this construction leaves one with a freedom to impose three
arbitrary conditions upon the ``constants" and/or their
derivatives. Out of convenience, the Lagrange constraint is often
imposed. It nullifies the convective term and thereby guarantees
that under perturbation the functional dependence of the velocity
upon the time and ``constants" stays the same as in the
undisturbed case. ``Constants" obeying this condition are called
osculating elements.

The ``constants" chosen to be canonical,
 are called
Delaunay elements, in the orbital case, or Andoyer elements, in
the spin case. (As some of the Andoyer elements are time-dependent
even in the free-spin case, the role of ``constants" is played by
these elements' initial values.) The Andoyer and Delaunay sets of
 elements share a feature not readily apparent: in certain
cases the standard equations render these elements non-osculating.

In orbital mechanics, elements calculated via the standard
planetary equations come out non-osculating when perturbations
depend on velocities.
To keep elements osculating under such perturbations, the
equations must be amended with extra terms that are
{\underline{not}} parts of the disturbing function (Efroimsky and
Goldreich 2003, 2004). For the Kepler elements, this merely
complicates the equations. In the case of Delaunay
parameterisation, these extra terms  not only complicate the
equations, but  also destroy their canonicity. So under
velocity-dependent disturbances, osculation and canonicity are
incompatible.

Similarly, in spin dynamics the Andoyer elements come out
non-osculating under angular-velocity-dependent perturbation (a
switch to a noninertial frame being one such case). Amendment of
the dynamical equations only with extra terms in the Hamiltonian
makes the equations render nonosculating Andoyer elements. To make
them osculating, more terms must enter the equations (and the
equations will no longer be canonical).

It is often convenient to deliberately deviate from osculation by
substituting the Lagrange constraint with an arbitrary condition
that gives birth to a family of nonosculating elements. The
freedom in choosing this condition is analogous to the gauge
freedom. Calculations in nonosculating variables are
mathematically valid and sometimes highly advantageous, but their
physical interpretation is nontrivial. For example, nonosculating
orbital elements parameterise instantaneous conics
{\underline{not}} tangent to the orbit, so the nonosculating
inclination will be different from the real inclination of the
physical orbit.

We present examples of situations in which ignoring of the gauge
freedom (and of the unwanted loss of osculation) leads to
oversights.
  }
  \end{abstract}
%
%
%
 \pagebreak


\section{Introduction}

\subsection{~~~Historical Prelude}

The orbital dynamics is based on the variation-of-parameters
method, invention whereof is attributed to Euler (1748, 1753) and
Lagrange (1778, 1783, 1808, 1809, 1810). Though both greatly
contributed to this approach, its initial sketch was offered circa
1687 by Newton in his unpublished {\it{Portsmouth Papers}}. Very
succinctly, Newton brought up this issue also in Cor. 3 and 4 of
Prop. 17 in the first book of his {\it{Principia}}.
 ~\\

Geometrically, the part and parcel of this method is
representation of an orbit as a set of points, each of which is
contributed by a member of some chosen family of curves
$C(\kappa)$, where $\kappa$ stands for a set of constants that
number a particular curve within the family. (For example, a set
of  three constants $\,\kappa\,=\,\{\,a,\,b,\,c\,\}\,$ defines one
particular hyperbola $\,y\,=\,ax^2\,+\,bx\,+\,c\,$ out of many).
This situation is depicted on Fig.1. Point $\,A\,$ of the orbit
coincides with some point $\lambda_1$ on a curve $C(\kappa_1)$.
Point $\,B\,$ of the orbit coincides with point $\lambda_2$ on
some other curve $C(\kappa_2)$ of the same family, etc. This way,
orbital motion from $\,A\,$ to $\,B\,$ becomes a superposition of
motion along $C_{\kappa}$ from $\lambda_1$ to $\lambda_2$ and a
gradual distortion of the curve $C_{\kappa}$ from the shape
$C(\kappa_1)$ to the shape $C(\kappa_2)$. In a loose language, the
motion along the orbit consists of steps along an instantaneous
curve $C(\kappa)$ which itself is evolving while those steps are
being made.
 \begin{center}
        \epsfxsize=105mm
        \epsfbox{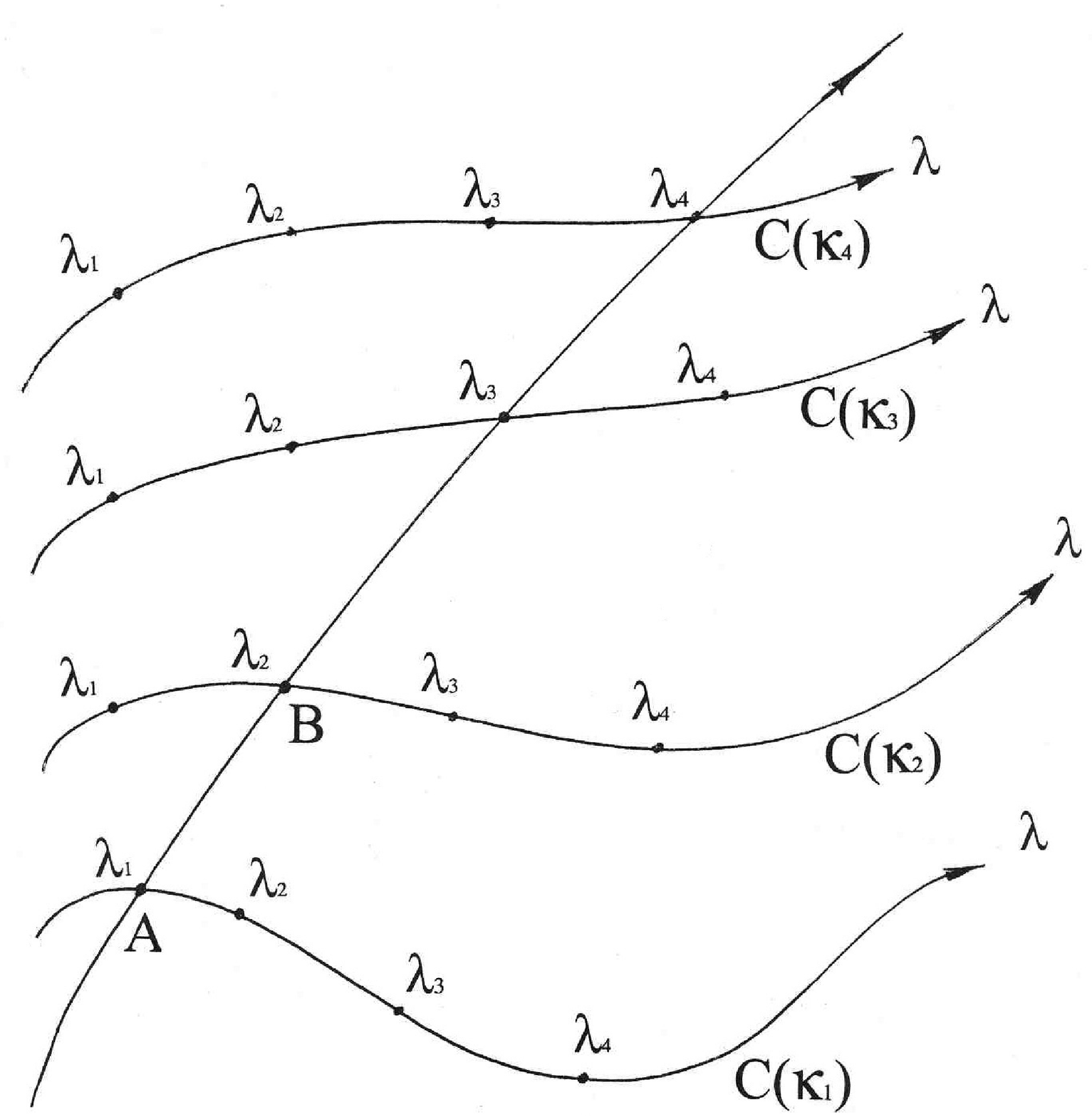}
 \end{center}
  \mbox{\small
 \parbox[b]{5.1in}{{\underline{Fig.1.}}~~~Each point of the orbit is contributed by a member
 of some family of curves $\,C(\kappa)\,$ of a certain type, $\,\kappa\,$ standing for a set
 of constants that number a particular curve within the family. Motion from A to B is,
 first, due to the motion along the curve $\,C(\kappa)\,$ from $\,\lambda_1\,$ to
 $\,\lambda_2\,$ and, second, due to the fact that during this motion the curve itself was
 evolving from $\,C(\kappa_1)\,$ to $\,C(\kappa_2)\,$.\\
  }}\\
Normally, the family of curves $C_{\kappa}$ is chosen to be that
of ellipses or that of hyperbolae, $\kappa$ being six orbital
elements, and $\lambda$ being the time. However, if we disembody
this idea of its customary implementation, we shall see that it is
of a far more general nature and contains three aspects:

{\underline{\textbf{1.}}} ~A trajectory may be assembled of points
contributed by a family of curves of an essentially arbitrary
type, not just conics.

{\underline{\textbf{2.}}} ~It is not necessary to choose the
family of curves tangent to the orbit. As we shall see below, it
is often beneficial to choose them nontangent. We shall also see
examples when in orbital calculations this loss of tangentiality
(loss of osculation) takes place and goes unnoticed.

{\underline{\textbf{3.}}} ~The approach is general and can be
applied, for example, to Euler's angles. A disturbed rotation can
be thought of as a series of steps (small turns) along different
Eulerian cones. An Eulerian cone is an orbit (on the Euler angles'
manifold) corresponding to an unperturbed spin state. Just as a
transition from one instantaneous Keplerian conic to another is
caused by disturbing forces, so a transition from one
instantaneous Eulerian cone to another is dictated by external
torques or other perturbations. Thus, in the attitude mechanics,
the Eulerian cones play the same role as the Keplerian conics do
in the orbital dynamics. Most importantly, a perturbed rotation
may be ``assembled" of the Eulerian cones in an osculating or in a
nonosculating manner. An unwanted loss of osculation in attitude
mechanics happens in the same way as in the theory of orbits, but
is much harder to notice. On the other hand, a deliberate choice
of nonosculating rotational elements in attitude mechanics may
sometimes be beneficial.
 ~\\

From the viewpoint of calculus, the concept of variation of
parameters looks as follows. We have a system of differential
equations to solve (``system in question") and a system of
differential equations (``fiducial system") whose solution is
known and contains arbitrary constants. We then use the known
solution to the fiducial system as an ansatz for solving the
system in question. The constants entering this ansatz are endowed
with time dependence of their own, and the subsequent substitution
of this known solution into the system in question yields
equations for the ``constants." The number of ``constants" often
exceeds that of equations in the system to solve. In this case we
impose, by hand, arbitrary constraints upon the ``constants." For
example, in the case of a reduced $\,N$-body problem, we begin
with $\,3(N\,-\,1)\,$ unconstrained second-order equations for
$\,3(N\,-\,1)\,$ Cartesian coordinates. After a change of
variables from the Cartesian coordinates to the orbital
parameters, we end up with $\,3(N\,-\,1)\,$ differential equations
for the $\,6(N\,-\,1)\,$ orbital variables. Evidently,
$\,3(N\,-\,1)\,$ constraints are necessary.\footnote{~In a fixed
Cartesian frame, any solution to the unperturbed reduced 2-body
problem can be written as
 \ba
 \nonumber
 x_j~=~f_j(t\,, \;C_1\,,\;.\;.\;.\;,\;C_6)~~~,~~~~~j\;=\;1\,,\;2\,,\;3~~~,
 \,~~~\\
 \nonumber
 \dot{x}_j\;=\;g_j(t\,,\;
 C_1\,,\;.\;.\;.\;,\;C_6)\;\;,\;\;\;\;\;g_j\;\equiv\;
 \left(\frac{\partial f_j}{\partial t}\right)_C~~~~~~~
 \ea
the adjustable constants $C$ standing for orbital elements. Under
disturbance, the solution is sought as
 \ba
 \nonumber
 x_j\;=\;f_j(t\,,
 \;C_1(t)\,,\;.\;.\;.\;,\;C_6(t)\,)~~~,~~~~~j\;=\;1\,,\;2\,,\;3\,~~~,
 ~~~~~~~~~~~~~~~~~~~~~~~~~~~~~~~~~~~~~~~~~~~~~~~~~~~~~~~~~~~~~~~~~~~~~~\\
 \nonumber
 \dot{x}_j\;=\;g_j(t\,,\;
 C_1(t)\,,\;.\;.\;.\;,\;C_6(t)\,)\;+\;
 \Phi_j(t\,,\;
 C_1(t)\,,\;.\;.\;.\;,\;C_6(t)\,)
 \;\;,\;\;\;\;\;g_j\;\equiv\;
 \left(\frac{\partial f_j}{\partial t}\right)_C\;\;,\;\;\;\;\;
 \Phi_j\;\equiv\;\sum_{r}\,\frac{\partial f_j}{\partial
 C_r}\;\dot{C}_r\;\;.
 \ea
 Insertion of
 $x_j=f_j(t,\,C)$ into the
 perturbed gravity law yields three scalar equations for six
 functions $C_r(t)$. This
 necessitates imposition of three conditions upon $C_r$
 and $\dot{C}_r$. Under the simplest choice $\Phi_j=0
 \,,\,\,j=1,2,3$,
 the perturbed physical velocity $\dot{x}_j(t,\, C)$ has the same functional form as the
 unperturbed
 $g_j(t,\,C)\,$. Therefore, the instantaneous
 conics become tangent to the orbit (and the orbital
 elements $\,C_r(t)\,$ are called osculating).}
 To this end, the so-called Lagrange constraint (the condition of the
 instantaneous conics being tangent to the physical orbit) is introduced almost
 by default, because it is regarded natural. Two things should be mentioned in
 this regard:

First, what seems natural is not always optimal. The freedom of
choice of the supplementary condition (the gauge freedom) gives
birth to an internal symmetry (the gauge symmetry) of the problem.
Most importantly, it can be exploited for simplifying the
equations of motion for the ``constants." On this issue we shall
dwell in the current paper.

Second, the entire scheme may, in principle, be reversed and used
to solve systems of differential equations with constraints.
Suppose we have $N+M$ variables $\;C_j(t)\;$ obeying a system of
$N$ differential equations of the second order and $M$ constraints
expressed with first-order differential equations or with
algebraic expressions. One possible approach to solving this
system will be to assume that the variables $\;C_j\;$ come about
as constants emerging in a solution to some fiducial system of
differential equations. Then our $N$ second-order differential
equations for $\;C_j(t)\;$ will be interpreted as a result of
substitution of such an ansatz into the fiducial system with some
perturbation, while our $M$ constraints will be interpreted as
weeding out of the redundant degrees of freedom. Unfortunately,
this subject is out of the scope of our paper, and
it will be developed somewhere else.\\

\subsection{~~~The simplest example of gauge freedom.}

Variation of constants first emerged in the nonlinear context of
celestial mechanics and later became a universal tool. We begin
with a simple example offered in Newman \& Efroimsky (2003)

A harmonic oscillator disturbed by a force $\Delta F(t)$ gives
birth to the initial-condition problem
 \be
 \ddot x \;+\; \;x\; =\;\Delta F(t)\;\;\;,\;\;\;\;\;{\mbox{with}}\;\;x(0)\;\;
 {\mbox{and}}\;\;\dot x (0)\;\;{\mbox{known}}\;\;,
 \label{1}
 \ee
whose solution may be sought using ansatz
 \be
 x = C_1(t) \sin t + C_2(t) \cos t \;\;\;.
 \label{2}
  \ee
 This will lead us to
  \be
  \dot x = \left[\;\dot C_1(t) \sin t
+ \dot C_2(t) \cos t \;\right]\;+\; C_1(t) \cos t - C_2(t) \sin t
\;\;\;.
  \label{3}
  \ee
It is common, at this point, to put the sum
$\;\left[\;{\dot{C}}_1(t) \sin t + {\dot{C}}_2(t) \cos
t\;\right]\;$ equal to zero, in order to remove the ambiguity
stemming from the fact that we have only one equation for two
variables. Imposition of this constraint is convenient but not
obligatory. A more general way of fixing the ambiguity may be
expressed as
 \be
 \dot C_1(t) \sin t + \dot C_2(t) \cos
t \,=\,\phi(t)\;\;,
 \label{4}
 \ee
$\phi(t)\,$ being an arbitrary function of time. This entails:
 \be
 \ddot x = \dot \phi\;+\;\dot C_1(t) \cos t - \dot C_2(t) \sin t - C_1(t)
 \sin t - C_2(t) \cos t\;\;\;,
 \label{5}
 \ee
 summation whereof with (\ref{2}) gives:
 \be
 \ddot x + x
= \dot \phi\;+\;\dot C_1(t) \cos t - \dot C_2(t) \sin t\;\;\;.
 \label{6}
 \ee
 Substitution thereof into (\ref{1}) yields
 the dynamical equation re-written in terms of the ``constants"
 $\,C_1\,,\,C_2\,$. This equation, together with identity (\ref{4}), will constitute the following system:
 \ba
 \nonumber
\dot \phi\;+\;\dot C_1(t) \cos t - \dot C_2(t) \sin t =\;\Delta F(t)\;\;\;,\\
 \label{7}
 \\
 \nonumber
  \dot C_1(t) \sin t + \dot C_2(t) \cos t \,=\,\phi(t)\;\;\;,\;\;\;\;\;\;\;
  \ea
This leads to
 \ba
 \nonumber
 \dot C_1\;=\;\,\Delta F\;\cos t\;-\;\frac{d}{dt} \,\left( \phi \cos t \right)
 \;\;\;\;\;\;\\
 \label{8}
 \\
 \nonumber
 \dot C_2\;=\,-\,\Delta F\;\sin t\;+\;\frac{d}{dt} \,\left( \phi
 \sin t \right)\;\;\,,
 \ea
the function $\;\phi(t)\;$ still remaining
arbitrary.\footnote{~Function $\;\phi(t)\;$ can afford being
arbitrary, no matter what the initial conditions are to be.
Indeed, for fixed $x(0)$ and $\dot x(0)$, the system $\;C_2(0) =
x(0)\;,\; \; \phi (0) + C_1(0) = \dot x(0)\;$ solves for $C_1(0)$
and $C_2(0)$ for an arbitrary choice of $\phi (0)$.} Integration
of (\ref{8}) entails:
 \ba
  \nonumber
C_1\;=\;\,\int^{t}\,\Delta F\;\cos t'\;dt'\;-\;\phi \cos t \; + \;a_1\;\;\;\\
 \label{9}
 \\
 \nonumber
 C_2\; = \,-\,\int^{t}\,\Delta F\;\sin t'\;dt'\; + \;\phi \sin t \; +
 \;a_2\;\;\,.
 \ea
 Substitution of (\ref{9}) into (\ref{2}) leads to complete
 cancellation of the $\;\phi\;$ terms:
 \ba
 x= C_1 \sin t + C_2 \cos t=\,-\,\cos t\,\int^t \Delta F\,\sin t'\,dt'\,+\,\sin t\,\int^t
 \Delta F\,\cos t' \, dt'\,+\,a_1\,\sin t\,+\,a_2\,\cos t\;\;\;\;\;
 \label{10}
  \ea
 Naturally, the physical trajectory $\,x(t)\,$ remains invariant under the choice of gauge function $\;
 \phi(t)\;$, even though the mathematical description (\ref{9}) of this motion in terms of the
 parameters $\,C$ is gauge dependent. It is, however, crucial that a numerical solution of
 the system (\ref{8}) will come out $\,\phi$-dependent, because the numerical error will be sensitive to
 the choice of $\,\phi(t)$. This issue is now being studied
 by P. Gurfil and I. Klein, and the results are to be published soon. (Gurfil \& Klein 2006)

 It remains to notice that (\ref{8}) is a simple analogue
 to the Lagrange-type system of planetary equations, system that,
 too, admits gauge freedom. (See subsection 2.2 below.)\\

 \subsection{Gauge freedom under a variation of the Lagrangian}

The above example permits an evident extension. (Efroimsky
2002a,b.) Suppose some mechanical system obeys the equation
 \ba
 \ddoterbold\;=\;\Fbold \left(\,t\,,\;\erbold\,,\;\doterbold\,\right)\;\;\;,\;\;\;\;
 \label{11}
 \ea
whose solution is known and has a functional form
 \ba
 \erbold\;=\;\efbold\left(\,t\,,\;C_1\,,\;.\;.\;.\;,\;C_6\,\right)\;\;\;,\;\;\;
 \label{12}
 \ea
$C_j\;$ being adjustable constants to vary only under disturbance.

When a perturbation $\;\Delta \Fbold\;$ gets switched on, the
system becomes:
 \ba
 \ddoterbold\;=\;\Fbold\left(\,t\,,\;\erbold\,,\;\doterbold\,\right)\;+
 \;\Delta \Fbold\left(\,t\,,\;\erbold\,,\;\doterbold\,\right)\;\;\;,\;\;\;\;
 \label{13}
 \ea
and its solution will be sought in the form of
 \ba
 \erbold\;=\;\efbold\left(\,t\,,\;C_1(t)\,,\;.\;.\;.\;,\;C_6(t)\,\right)\;\;\;.\;\;\;
 \label{14}
 \ea
Evidently,
 \ba
 \doterbold\;=\;\frac{\partial \efbold}{\partial
 t}\;+\;\Phibold\;\;\;,\;\;\;\;\;\;
 \Phibold\;\equiv\;\sum_{j=1}^{6}\,\frac{\partial \efbold}{\partial
 C_j}\;\dot{C}_j\;\;.\;\;\;
 \label{15}
 \ea
In defiance of what the textbooks advise, we do
{\underline{\textbf{not}}}$\,$ put $\Phibold$ nil. Instead, we
proceed further to
 \ba
 \ddoterbold\;=\;\frac{\partial^2 \efbold}{\partial t^2}\;+\;
 \sum_{j=1}^{6}\,\frac{\partial^2 \efbold}{\partial t \;\partial
 C_j}\;\dot{C}_j\; +\;\dot{\Phibold}\;\;\;,\;\;\;
 \label{16}
 \ea
dot standing for a {\underline{full}} time derivative. If we now
insert the latter into the perturbed equation of motion (\ref{13})
and if we recall that, according to (\ref{11}),\footnote{~We
remind that in (\ref{11}) there was no difference between a
partial and a full time derivative, because at that point the
integration ``constants" $\,C_i$ were indeed constant. Later, they
acquired time dependence, and therefore the full time derivative
implied in (\ref{15} - \ref{16}) became different from the partial
one implied in (\ref{11}).} $\;\partial^2 \efbold/\partial
t^2\;=\;\Fbold\;$, then we shall obtain the equation of motion for
the new variables $\;C_j(t)\;$:
 \ba
 \sum_{j=1}^{6}\,\frac{\partial^2 \efbold}{\partial t \;\partial C_j}\;\dot{C}_j\;
 +\;\dot{\Phibold}\;=\;\Delta \Fbold
 \label{17}
 \ea
 where
 \ba
 \Phibold\;\equiv\;\sum_{j=1}^{6}\,\frac{\partial \efbold}{\partial
 C_j}\;\dot{C}_j\;\;\;\;\;
 \label{18}
 \ea
so far is merely an identity. It will become a constraint after we
choose a particular functional form
$\,\Phibold\left(t\,;\;C_1\,,\;.\;.\;.\;,\;C_6\right)\,$ for the
gauge function $\,\Phibold\,$, i.e., if we choose that the sum
$\;\textstyle{\sum\,\frac{\textstyle\partial
\efbold}{\textstyle\partial C_j}\;\dot{C}_j}\;$ be equal to some
arbitrarily fixed function $\, \Phibold
(t\,;\,C_1\,,\,.\,.\,.\,,\,C_6)\,$ of the time and of the variable
``constants." This arbitrariness exactly parallels the gauge
invariance in electrodynamics: on the one hand, the choice of the
functional form of $\,\Phibold (t\,;\,C_1\,,\,.\,.\,.\,,\,C_6)\,$
will never\footnote{~Our usage of words ``arbitrary" and ``never"
should be limited to the situations where the chosen gauge
(\ref{21}) does not contradict the equations of motion (\ref{20}).
This restriction, too, parallels a similar one present in field
theories. Below we shall encounter a situation where this
restriction becomes crucial.} influence the eventual solution for
the physical variable $\,\erbold\,$; on the other hand, though, a
qualified choice may considerably simplify the process of finding
the solution. To illustrate this, let us denote by
$\,{\bf{
{g}}}(t\,,\;C_1\,,\;.\;.\;.\;,\;C_6)\,$ the functional dependence
of the unperturbed velocity on the time and adjustable constants:
 \ba
 {\bf{
 g}}(t\,,\;C_1\,,\;.\;.\;.\;,\;C_6)\;\equiv\;\frac{\partial }{\partial
 t}\,\efbold (t\,,\;C_1\,,\;.\;.\;.\;,\;C_6)\;\;\;,
 \label{19}
 \ea
and rewrite the above system as
 \ba
 \sum_{j}\,\frac{\partial {\bf
 {g}}}{\partial C_j}\;\dot{C}_j\;=\;-\;\dot{\Phibold}
 \;+\;\Delta \Fbold
  \label{20}
 \ea
 \ba
 \sum_{j}\,\frac{\partial \efbold}{\partial C_j}\;\dot{C}_j\;=\;\Phibold\;\;.\;\;\;
 \label{21}
 \ea
If we now dot-multiply the first equation with $\;\partial \efbold
/\partial C_i\;$ and the second one with $\;\partial {\bf
g}/\partial C_i\;$, and then take the difference of the outcomes,
we shall arrive at
 \ba
 \sum_{j}\,\left[\,C_n\;C_j\,\right]\,\dot{C}_j\;=\;\left(\,\Delta \Fbold\;-\;\dot{\Phibold}\,
 \right)\,\cdot\,\frac{\partial \efbold}{\partial C_n}\;-\;\Phibold\,\cdot\,
 \frac{\partial {\bf
 {g}}}{\partial C_n}~\;\;,
 \label{22}
 \ea
 the Lagrange brackets being defined in a gauge-invariant (i.e., $\Phibold$-independent)
 fashion.\footnote{~The Lagrange-bracket matrix is defined in a gauge-invariant way:
 \ba
 \nonumber
  \sum_{j}\,\left[\,C_n\;C_j\,\right]\;\equiv\;
 \frac{\partial \efbold}{\partial C_n}\,\cdot\,\frac{\partial {\bf{
 {g}}} }{\partial C_j}\;-\;
 \frac{\partial {\bf{\vec{g}}} }{\partial C_n}\,\cdot\,\frac{\partial \efbold}{\partial
 C_j}\;\;\;.
 \ea
 and so is its inverse, the matrix composed of
 the Poisson brackets
  \ba
  \nonumber
 \left\{\,C_n\;C_j\,\right\}\;\equiv\;
 \frac{\partial C_n}{\partial \efbold}\,\cdot\,\frac{\partial C_j}{\partial {\bf
 {g}}}\;-\;
 \frac{\partial C_n}{\partial {\bf
 {g}}}\,\cdot\,\frac{\partial C_j}{\partial
 \efbold}\;\;\;.
 \ea
 Evidently, (\ref{22}) yields
  \ba
  \nonumber
  \dot{C}_n\;=\;\sum_j\left\{\,C_n\;C_j\,\right\}\;
  \left[\,\frac{\partial \efbold}{\partial C_j}\,\cdot\,\left(\,\Delta \Fbold\;-\;\dot{\Phibold}
   \,\right)
  \;-\;\Phibold\,\cdot\,\frac{\partial {\bf
  {g}}}{\partial C_j}  \,\right]\;\;\;.
  \ea
  }
 If we agree that $\Phibold$ is a function of both the time and the parameters $\,C_n\,$,
 but not of their derivatives,\footnote{~The necessity to fix a functional form of $\;
  \Phibold(\,t\,;\;C_1\,,\;.\;.\;.\;,\;C_6\,)\;$, i.e., to impose three
arbitrary conditions upon the ``constants" $\;C_j\;$, evidently
follows from the fact that, on the one hand, in the ansatz
(\ref{14})
 we have six variables $\;C_n(t)\;$ and, on the other hand, the number of scalar equations
 of motion (i.e., Cartesian projections of the perturbed vector equation
(\ref{13})$\,$) is only three. This necessity will become even
more
 mathematically transparent after we cast the perturbed equation (\ref{13}) into
the normal form of Cauchy. (See Appendix.)} then the right-hand
side of (\ref{22}) will implicitly
 contain the first time derivatives of $\,C_n\,$. It will then be reasonable to move these
 to the left-hand side. Hence, (\ref{22}) will be reshaped into
 \be
 \sum_j\;\left(\,[C_n\;C_j]\;+\;\frac{\partial \fbold
 }{\partial C_n}\,\cdot\,\frac{\partial \Phibold}{\partial C_j}\;\,\right)\,\frac{dC_j}{dt}\;=\;
\frac{\partial \bf {
 f}}{\partial C_n}\,\cdot\, {\Delta {\Fbold}}\;-\;
\frac{\partial{\bf {
 f}}}{\partial C_n}\,\cdot\,\frac{\partial \bf {
 \Phi}}{\partial t}\;-\;\frac{\partial \bf {
 g}}{\partial C_n} \,\cdot\,\Phibold \;\;\;\;.
 \label{23}
 \ee
This is the general form of the gauge-invariant perturbation
equations, that follows from the variation-of-parameters method
applied to problem (\ref{13}), for an arbitrary perturbation
$\;\Fbold(\erbold,\,{\bf{\dot{\erbold}}},\,t)\,$ and under the
simplifying assumption that the arbitrary gauge function
$\;\Phibold\;$ is chosen to depend on the time and the parameters
$\;C_n\;$, but not on their derivatives.\footnote{~We may also
impart the gauge function with dependence upon the parameters'
time derivatives of all orders. This will yield
higher-than-first-order derivatives in equation (\ref{23}). In
order to close this system, one will then have to impose
additional initial conditions, beyond those on $\;\erbold\;$ and
$\;\doterbold\;$.} Assume that our problem (\ref{13}) is not
simply mathematical but is an equation of motion for some physical
setting, so that $\Fbold $ is a physical force corresponding to
some undisturbed Lagrangian $\,{\cal L}_o\,$, and $\,\Delta
\Fbold\,$ is a force perturbation generated by a Lagrangian
variation $\,\Delta{\cal L}\,$. If, for example, we begin with
 $\,{\cal L}_o(\erbold\,,\,\doterbold\,,\,t)\,=\,\doterbold^{\left.\,\right. 2}/2\,-
 \,U({\erbold\,,\,t})\,$, momentum $\,{\bf
 {p}}\,=\,\doterbold\,$, and Hamiltonian
 $\,{\cal H}_o(\erbold\,,\,{\bf
 {p}}\,,\,t)\,=\,{\bf
 p}^2/2\,+\,U(\erbold\,,\,t)\,$,
 then their disturbed counterparts will read:
 %
 \ba
 {\cal L}(\erbold\,,\;\doterbold\,,\;t)~=~\frac{\doterbold^{\left.\,\right. 2}}{2}
 \;-\;U(\erbold)\;+\;\Delta {\cal L} ( \erbold,
 \,\doterbold ,\,t) \;\;\;,
 \label{24}
 \ea
 \ba
 {\pbold}\;=\;\doterbold\;+\;\frac{\partial \Delta {\cal L}}{\partial
 \doterbold}\;\;\;,~~~~~~~~~~~~~~~~~~~~~~~~~~~~~~~~~
 \label{25}
 \ea
 \ba
 {\cal H}\;=\;{\bf
 p}\;\doterbold\;-\;{\cal L}\;=\;
 \frac{{\bf
 p}^{\left.
\,\right. 2}}{2}\;+ \;U\;+\;\Delta
 {\cal H}\;\;\;,~~~~~
 \label{26}
 \ea
 \ba
 \Delta {\cal H}\;\equiv\;-\;\Delta
 {\cal L}\;-\;\frac{1}{2}\,\left(\frac{\partial \,\Delta {\cal L}}{\partial \doterbold }
 \right)^2\;\;.~~~~~~~~~~~~~~
 \label{27}
 \ea
The Euler-Lagrange equations written for the perturbed Lagrangian
(\ref{24}) are:
 \be
 {\bf{\ddot {
 r}}}\;=\;-\;\frac{\partial U}{\partial \erbold} \;+\;
\Delta
 {\Fbold}\;\;\;\;,
 \label{28}
 \ee
where the disturbing force is given by
 \be
 \Delta {\Fbold}\;\equiv\;\frac{\partial \,\Delta {\cal L}}{\partial
 \erbold}\;-\;\frac{d}{dt}\,\left(\frac{\partial \,\Delta {\cal L}}{\partial
 \doterbold}\right)\;\;\;\;.
 \label{29}
 \ee
Its substitution in (\ref{23}) yields the generic form of the
equations in terms of the Lagrangian disturbance (Efroimsky \&
Goldreich 2004):
  \begin{eqnarray}
 \nonumber
\sum_j\;\left(\;[C_n\;C_j]\;+
 \;\frac{\partial {\efbold }}{\partial C_n}\,\cdot\,
  \frac{\partial }{\partial C_j}\;
  \left(\frac{\partial \,\Delta \cal
 L}{\partial \doterbold}\;+\;{\Phibold}
  \right)\;\right)
  \frac{dC_j }{dt }\;\;=
~~~~~~~~~~~~~~~~~~~~~~~~~~~~~~~~~~~~~~~~~\\
 \label{30}\\
 \nonumber
 \frac{\partial }{\partial C_n}\,\left[\Delta {\cal L}\,+\,\frac{1}{2}\,
 \left(\frac{\partial \,\Delta \cal L}{\partial \doterbold} \right)^2 \right]\;-\;
 \left( \frac{\partial \bf
 g}{\partial C_n}\;+\;\frac{\partial \efbold }{\partial C_n}\;
 \frac{\partial}{\partial t}\;+\;\frac{\partial \,\Delta \cal L}{\partial \doterbold}\;
 \frac{\partial }{\partial C_n} \right)\cdot\left(\Phibold\,+\,\frac{\partial \,
 \Delta \cal L}{\partial \doterbold}
 \right)
 \;\;\;.\;\;
  \end{eqnarray}
This equation not only reveals the convenience of the special
gauge
  \begin{eqnarray}
 \Phibold\;=\;-\;\frac{\partial\,\Delta \cal L}{\partial
 \doterbold}\;\;\;,
 \label{31}
  \end{eqnarray}
 (which reduces to $~\Phibold\,=\,0~$ in the case of velocity-independent perturbations), but
 also explicitly demonstrates how the Hamiltonian variation comes into play: it is easy to
 notice that, according to (\ref{27}), the sum in square brackets on the right-hand side of
 (\ref{30}) is equal to
 $~-\,\Delta{\cal H}~$, so the above equation takes the form
 $\;\,\sum_j\;[C_n\;C_j]\;\dot{C}_j\;=\;-\;{\partial \Delta {\cal{H}}}/{ \partial C_n }\;\,$.
 All in all, it becomes clear that the trivial gauge, $\;\Phibold\;=\;0\;$, leads to the
 maximal simplification of the variation-of-parameters equations expressed
 through the disturbing force: it follows from (\ref{22}) that
 \ba
 \sum_{j}\,\left[\,C_n\;C_j\,\right]\,\dot{C}_j\;=\;\Delta
 \Fbold\;\cdot\;
 \frac{\partial \efbold}{\partial C_n}\;\;\;,\;\;\;\;
 \mbox{provided~we~have~chosen}\;\;\Phibold\;=\;0\;\;\;.~~~~~~~~~~~
 \label{32}
 \ea
 However, the choice of the special gauge (\ref{31}) entails the
 maximal simplification of the variation-of-parameters equations
 when they are formulated via a variation of the Hamiltonian:
 \ba
 \sum_j\;[C_n\;C_j]\;\frac{dC_j }{dt }\;=\;-\;\frac{\partial \Delta{\cal{H}}}{\partial C_n}
 \;\;\;,\;\;\;\;\mbox{provided~we~have~chosen}\;\;\Phibold\;=\;
 -\;\frac{\partial\,\Delta \cal L}{\partial \doterbold}\;\;\;.\;\;\;\;
 \label{33}
 \ea
 It remains to spell out the already obvious fact that, in case the
 unperturbed force $\;\Fbold\;$ is given by the Newton gravity law
 (i.e., when the undisturbed setting is the reduced two-body problem),
 then the variable ``constants" $\;C_n\;$ are merely the orbital
 elements parameterising a sequence of instantaneous conics out of
 which we ``assemble" the perturbed trajectory through (\ref{14}). When the conics'
 parameterisation is chosen to be via the Kepler or the Delaunay
 variables, then (\ref{30}) yields the gauge-invariant version of
 the Lagrange-type or the Delaunay-type planetary equations, accordingly.
 Similarly, (\ref{22}) implements the gauge-invariant generalisation of the
 planetary equations in the Euler-Gauss form.

 From (\ref{22}) we see that the Euler-Gauss-type planetary equations will
 always assume their simplest form (\ref{32}) under the gauge choice
 $\;\Phibold\;=\;0\;$.
 In astronomy this choice is called ``the Lagrange constraint." It
 makes the orbital elements osculating, i.e., guarantees that the
 instantaneous conics, parameterised by these elements, are
 tangent to the perturbed orbit.

 From (\ref{33}) one can easily notice that the Lagrange- and
 Delaunay-type planetary equations simplify maximally under the
 condition (\ref{31}). This condition coincides with the Lagrange
 constraint $\;\Phibold\;=\;0\;$ when the perturbation depends
 only upon positions (not upon velocities or momenta). Otherwise,
 condition (\ref{31}) deviates from that of Lagrange, and the orbital elements
 rendered by equation (\ref{33}) are no longer osculating (so that the
 corresponding instantaneous conics are no longer tangent to the physical trajectory).

 Of an even greater importance will be the following observation.
 If we have a velocity-dependent perturbing force, we can always
 find the appropriate Lagrangian variation and, therefrom, the
 corresponding variation of the Hamiltonian. If now we simply add
 the negative of this Hamiltonian variation to the disturbing
 function, then the resulting equations (\ref{33}) will render not
 the osculating elements but orbital elements of a different type,
 ones satisfying the non-Lagrange constraint (\ref{31}). Since the
 instantaneous conics, parameterised by such non-osculating
 elements, will not be tangent to the orbit, then physical
 interpretation of such elements may be nontrivial. Besides, they
 will return a velocity different from the physical one.\footnote{~ We mean that
 substitution of the values of these elements in $\;{\bf
 g}(t\,;\;C_1(t)\,,\;
 .\;.\;.\;,\;C_6(t))\;$ will not give the right velocity. The correct physical
 velocity will be given by $\;\doterbold\;=\;{\bf
 g}\;+\;\Phibold\;$.} This pitfall is
 well camouflaged and is easy to fall in.

 These and other celestial-mechanics applications of the gauge freedom will be considered in
 detail in section 2 below.

\subsection{Canonicity versus osculation}

 One more relevant development will come from the theory of canonical perturbations. Suppose
 that in the absence of disturbances we start out with a system
  \ba
 \dot{q}\;=\;\frac{\partial {\cal H}^{(o)}}{\partial p}\;\;\;,\;\;\;\;\;~~
 \dot{p}\;=\;-\;\frac{\partial {\cal H}^{(o)}}{\partial q}
 ~~~.~~~~~~~~~~~~~~~~~~~~~~~~~~~~~
 \label{34}
 \ea
 $q\;$ and $\;p\;$ being the Cartesian or polar coordinates and their conjugated
 momenta, in the orbital case, or the Euler angles and their momenta, in the
 rotation case. Then we switch, via a canonical transformation
 \ba
 q\;=\;f(Q\,,\;P\,,\;t)\;\;\;,\;\;\;\;\;p\;=\;\chi(Q\,,\;P\,,\;t)\;~~~~~~~~~~~~~~~~~~~~~
 \label{35}
 \ea
 to
 \ba
 \dot{Q}\;=\;
 \frac{\partial {\cal H}^*}{\partial P}\;=0\;\;\;,\;\;\;\;\;
 \dot{P}\;=\;-\;\frac{\partial {\cal H}^*}{\partial Q}\;=\;0
 \;\;\;,\;\;\;\;
 {\cal H}^*\;=\;0\;\;,
 \label{36}
 \ea
 where $\;Q\;$ and $\;P\;$ denote the set of Delaunay elements, in the orbital
 case, or the initial values of the Andoyer variables, in the case of
 rigid-body rotation.

 This scheme relies on the fact that, for an unperturbed motion (i.e., for an unperturbed
 Keplerian conic, in an orbital case; or for an undisturbed Eulerian cone, in the spin case)
 a six-constant parameterisation may be chosen so that:\\
 ~\\
 ~~ \textbf{\underline{1.}}~~~the parameters are constants and, at the same time,
 are canonical variables $\,\{\,Q\,,\,P\,\}\,$ with a zero Hamiltonian
 $\,{\cal H}^*(Q,\,P)\,=\,0\,$; \\
 ~\\
 ~~ \textbf{\underline{2.}}~~$\,$for constant $\,Q\,$ and $\,P\,$, the
 transformation equations (\ref{35}) are mathematically equivalent to the
 dynamical equations (\ref{34}).\\

 \noindent
 Under perturbation, the ``constants" $Q,\,P$ begin
to evolve so that, after their substitution into
 \ba
 q\;=\;f\left(\,Q(t)\,,\;P(t)\,,\;t\,\right)\;\;\;,\;\;\;\;\;p\;=\;\chi(\,Q(t)\,,\;
 P(t)\,,\;t\,)\;\;\;, ~~~~~~~~~~~~~~
 \label{37}
 \ea
($f,\chi$ being the same functions as in (\ref{35})$\,$), the
resulting motion obeys the disturbed equations
  \ba
 \dot{q}\;=\;\frac{\partial \left({\cal H}^{(o)}\,+\,
 \Delta {\cal H}\right)}{\partial p}\;\;\;,\;\;\;\;\;~~
 \dot{p}\;=\;-\;\frac{\partial \left({\cal H}^{(o)}\,+\,
 \Delta {\cal H} \right)}{\partial q}
 ~~~.~~~~~~~~~~~~~~~
 \label{38}
 \ea
We also  want our ``constants" $\;Q\;$ and $\;P\;$ to remain
canonical and to obey
  \ba
 \dot{Q}\;=\;\frac{\partial \left({\cal H}^*\,+\,
 \Delta {\cal H}^* \right)}{\partial P}\;\;\;,\;\;\;\;\;~~
 \dot{P}\;=\;-\;\frac{\partial \left({\cal H}^*\,+\,
 \Delta {\cal H}^* \right)}{\partial Q}~~~~~~~~~~~~~~~~~~
 \label{39}
 \ea
 where
 \ba
 {\cal H}^*\,=\;0\;\;\;\;\mbox{and}\;\;\;\;\;\Delta {\cal
 H}^*\left(Q\,,\;P\,\;t\right)\;=\;\Delta {\cal
 H}\left(\,q(Q,P,t)\,,\;p(Q,P,t)\,,\;t\,\right)\;\;\;.
 \label{40}
 \ea
Above all, we wish the perturbed ``constants" $\,C\,=\,Q,\,P\,$
(the Delaunay elements, in the orbital case; or the initial values
of the Andoyer elements, in the spin case) to osculate. This means
that we want the perturbed velocity to be expressed by the same
function of $\,C_j(t)\,$ and $\,t\,$ as the unperturbed velocity.
Let us check when this is possible. The perturbed velocity is
 \ba
 \dot{q}\;=\;g\;+\;\Phi ~~~~~~~~~~~~~~~~~~~~~~~~~~~~~~
 \label{41}
 \ea
where
 \ba
 g(C(t),\,t)\;\equiv\;\frac{\partial q(C(t),\,t)}{\partial t}
 \;\;~~~~~~~~~~~~
 \label{42}
 \ea
is the functional expression for the unperturbed velocity, while
 \ba
 \Phi(C(t),\,t)\;\equiv\;\sum_{j=1}^6\,\frac{\partial q(C(t),\,t)}{\partial
 C_j}\;\dot{C}_j(t)\;
 \label{43}
 \ea
is the convective term. Since we chose the ``constants" $\,C_j\,$
to make canonical pairs $\,(Q,\,P)\,$ obeying (\ref{39} -
\ref{40}), then insertion of (\ref{39}) into (\ref{43}) will
result in
 \ba
 \Phi\;=\;\sum_{n=1}^3\,\frac{\partial q}{\partial
 Q_n}\;\dot{Q}_n(t)\;+\;\sum_{n=1}^3\,\frac{\partial q}{\partial
 P_n}\;\dot{P}_n(t)\;=\;\frac{\partial \Delta {\cal H}(q,\,p)}{\partial
 p}\;\;\;.
 \label{44}
 \ea
So canonicity is incompatible with osculation when $\Delta{\cal
H}$ depends on $p$. Our desire to keep the perturbed equations
(\ref{39}) canonical makes the orbital elements $\,Q\,,\,P\,$
nonosculating in a particular manner prescribed by (\ref{44}).
This breaking of gauge invariance reveals that the canonical
description is marked with ``gauge stiffness" (term suggested by
Peter Goldreich).

We see that, under a momentum-dependent perturbation, we still can
use the ansatz (\ref{37}) for calculation of the coordinates and
momenta, but can no longer use $\;\dot{q}\,=\,{\partial
q}/{\partial t}\;$ for calculating the velocities. Instead, we
must use $\;\dot{q}\,=\,{\partial q}/{\partial
t}\,+\,\partial\Delta {\cal H}/\partial p\,$, and the elements
$\,C_j\,$ will no longer be osculating. In the case of orbital
motion (when $\,C_j\,$ are the nonosculating Delaunay elements),
this will mean that the instantaneous ellipses or hyperbolae
parameterised by these elements will not be tangent to the orbit.
(Efroimsky \& Goldreich 2003.) In the case of spin, the situation
will be similar, except that, instead of an instantaneous
Keplerian conic, one will be dealing with an instantaneous
Eulerian cone -- a set of trajectories on the Euler-angles
manifold, each of which corresponds to some non-perturbed spin
state. (Efroimsky 2004.)

The main conclusion to be derived from this example is the
following: whenever we encounter a disturbance that depends not
only upon positions but also upon velocities or momenta,
implementation of the afore described canonical-perturbation
method necessarily yields equations that render nonosculating
canonical elements. It is possible to keep the elements
osculating, but only at the cost of sacrificing canonicity. For
example, under velocity-dependent orbital perturbations (like
inertial forces, or atmospheric drag, or relativistic correction)
the equations for osculating Delaunay elements will no longer be
Hamiltonian (Efroimsky 2002a,b).

Above in this subsection we discussed the disturbed velocity
$\,\dot{q}\,$. How about the disturbed momentum? For sufficiently
simple unperturbed Hamiltonians, it can be written down very
easily. For example, for $\;{\cal H}\;=\;{\cal H}_o\;+\;\Delta
{\cal H}\;=\;p^2/2m\;+\;U(q)\;+\;\Delta {\cal H}\;$ we get:
 \ba
 {{p}}\;=\;\dot{q}\;+\;\frac{\partial \Delta{\cal L}}{\partial
 \dot{q}}\;=\;g\;+\;\Phi\;+\;\frac{\partial \Delta{\cal L}}{\partial
 \dot{q}}\;=\;g\;+\;\left(\,\Phi\;-\;\frac{\partial \Delta {\cal H}}{\partial
 \dot{q}}\,\right)\;=\;g\;\;.~~~
 \label{45}
 \ea
In this case, the perturbed momentum $p$ coincides with the
unperturbed one, $g$. In application to the orbital motion, this
means that contact elements (i.e., the nonosculating orbital
elements obeying (\ref{31})$\,$), when substituted in
 $g(t\,;\,C_1\,,\,.\,.\,.\,,\,C_6)$, furnish not the correct
 perturbed velocity but the correct perturbed momentum, i.e., they
  osculate the orbit {\it{in phase space}}.
  Existence of such elements was pointed out long ago by Goldreich (1965)
  and Brumberg et al (1971).


\pagebreak

\section{{Gauge freedom in the theory of orbits.}}

\subsection{Geometrical meaning of the arbitrary gauge function $\;\Phibold$}

As explained above, the content of subsection 1.3 becomes merely a
formulation of the Lagrange theory of orbits, provided
$\;\Fbold\;$ stands for the Newton gravity force, so that the
undisturbed setting is the two-body problem. Then (\ref{22})
expresses the gauge-invariant (i.e., taken with an arbitrary gauge
$\,\Phibold(t\,;\,C_1\,,\,.\,.\,.\,,\,C_6)\,$) planetary equations
in the Euler-Gauss form. These equations render orbital elements
that are, generally, not osculating. Equation (\ref{32}) stands
for the customary Euler-Gauss-type system for osculating (i.e.,
obeying $\,\Phibold\,=\,0\,$) orbital elements.

Similarly, equation (\ref{30}) stands for the gauge-invariant
Lagrange-type or Delaunay-type (dependent upon whether $\;C_i\;$
stand for the Kepler or Delaunay variables) equations. Such
equations yield elements, which, generally, are not osculating. In
those equations, one could fix the gauge by putting
$\;\Phibold\;=\;0\;$, thus making the resulting orbital elements
osculating. However, this would be advantageous only in the case
of velocity-independent $\;\Delta {\cal L}\;$. Otherwise, a
maximal simplification is achieved through a deliberate refusal
from osculation: by choosing the gauge as (\ref{31}) one ends up
with simple equations (\ref{33}). Thus, gauge (\ref{31})
simplifies the planetary equations. (See equations (\ref{222} -
\ref{Delaunay.6}) below.) Besides, in the case when the Delaunay
parameterisation is employed, this gauge makes the equations for
the Delaunay variables canonical for reasons explained above in
subsection 1.4.

 The geometrical meaning of the convective term $\,\Phibold\,$ becomes evident if we recall that
 a perturbed orbit is assembled of points, each of which is donated by one
 representative of a sequence
 of conics, as on Fig.2 and Fig.3 where the ``walk" over the
 instantaneous conics may be undertaken either in {\underline{\textbf{a}}} non-osculating
 manner
 or in {\underline{\textbf{the}}} osculating manner. The physical velocity $\,\doterbold\,$
 is
 always tangent to the perturbed orbit, while the unperturbed Keplerian velocity
 $\,{\bf{
 g}}\,\equiv\,{\partial \efbold}/{\partial t}\,$ is tangent to the
 instantaneous conic.
 Their difference is the convective term $\,\Phibold$.
 So if we use non-osculating orbital elements, then
 insertion of their values in $\,\efbold(t\,;\,C_1\,,\,.\,.\,.\,,\,C_6)\,$
 will yield a correct position of the body. However, their
 insertion in
 $\,{\bf{
 {g}}}(t\,;\,C_1\,,\,.\,.\,.\,,\,C_6)\,$ will NOT
 give the right velocity. To get the correct velocity, one will
 have to add $\,\Phibold\,$. (See Appendix for a more formal mathematical treatment
 in the normal form of Cauchy.)

 When using non-osculating orbital elements, we must always be careful about their physical
 interpretation. On Fig. 2, the instantaneous conics are not supposed to be tangent to the orbit,
 nor are they supposed to be even coplanar thereto. (They may be even perpendicular to the orbit!
 -- why not?) This means that, for example, the non-osculating element $\;\inc\;$ may
 considerably differ from the real, physical inclination of the orbit.

We would add that the arbitrariness of choice of the function
$\,\Phibold(t\,,\,C_1(t)\,,\,.\,.\,.\,,\,C_6(t))\,$ had been long
known but never used in astronomy until a recent effort undertaken
by several authors (Efroimsky 2002a,b ; Newman \& Efroimsky 2003;
Slabinski 2003; Efroimsky \& Goldreich 2003, 2004; Gurfil 2004;
Efroimsky 2005c) Substitution of the Lagrange constraint
$\,\Phibold\,=\,0\,$ with alternative choices does not influence
the physical motion, but alters its mathematical description
(i.e., renders different values of the orbital parameters
$\,C_i(t)$). Such invariance of a physical picture under a change
of parameterisation goes under the name of gauge freedom. It is a
part and parcel of electrodynamics and other field theories. In
mathematics, it is described in terms of fiber bundles. A clever
choice of gauge often simplifies solution of the equations of
motion. On the other hand, the gauge invariance may have
implications upon numerical procedures. We mean the so-called
"gauge drift," i.e., unwanted displacement in the gauge function
$\,\Phibold$, caused by accumulation of numerical errors in the
constants.
 \begin{center}
        \epsfxsize=56mm
        \epsfbox{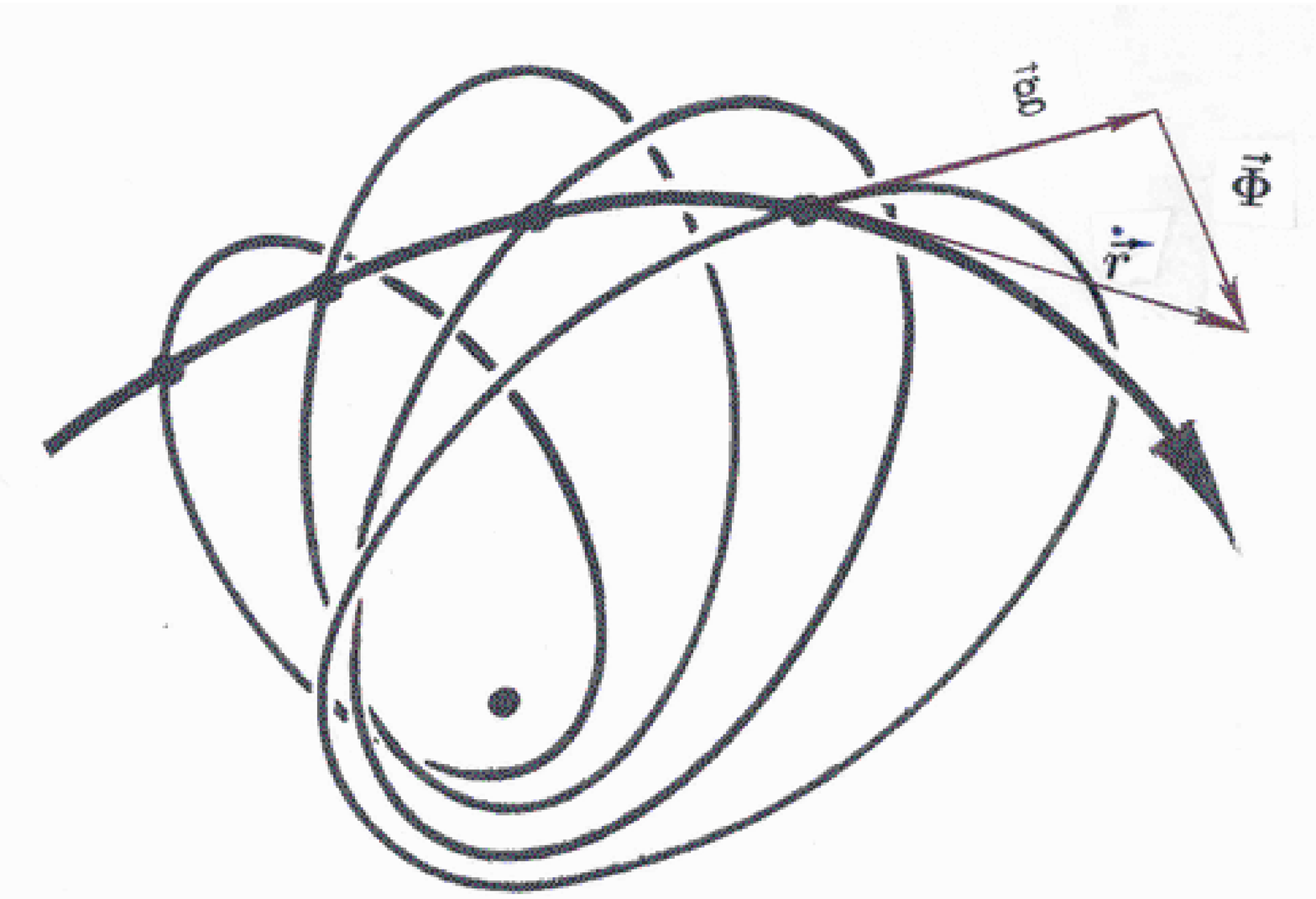}
  \end{center}
  \mbox{\small
 \parbox[b]{5.1in}{{\underline{Fig.2.}} ~~ \small The orbit
 is a set of points, each of which is donated by one of the
 confocal instantaneous ellipses that are \textbf{not} supposed to be tangent or even coplanar
 to the orbit. As a result, the
physical velocity $\doterbold$ (tangent to the orbit) differs from
the unperturbed Keplerian velocity $\bf
 g$ (tangent to the
ellipse). To parameterise the depicted sequence of non-osculating
ellipses, and to single it out of the other sequences, it is
suitable to employ the difference between $\doterbold $ and
$\bf
 g $, expressed as a function of time and six
(non-osculating) orbital elements:
 $\,
 \Phibold(t\,,\,C_1\,,\,.\,.\,.\,,\,C_6)\,=\,
 \doterbold(t\,,\,C_1\,,\,.\,.\,.\,,\,C_6)\,-\,
 {\bf
 g}(t\,,\,C_1\,,\,.\,.\,.\,,\,C_6)\;.\;
 $
 Evidently,
 \ba
 \nonumber
 \doterbold\,=\,\frac{\partial \erbold}{\partial
 t}\,+\,\sum_{j=1}^{6}\frac{\partial C_j}{\partial
 t}\;\dot{C}_j\;=\;{\bf
 g}\;+\;\Phibold\;\;\;,
 \ea
 where the unperturbed Keplerian velocity is ${\bf\vec g} \equiv {\partial \erbold}/
 {\partial t}$. The convective term, which emerges under perturbation, is $\Phibold
 \equiv \sum \left(\partial \erbold/\partial C_j \right)\dot{C}_j$. When a particular
 functional dependence of $ \Phibold $ on time and the elements is fixed, this function,
 $\Phibold(t\,,\,C_1\,,\,.\,.\,.\,,\,C_6)$, is called gauge function or gauge velocity or,
 simply, gauge.}}

 ~\\

 ~\\

 ~\\

 \begin{center}
        \epsfxsize=55mm
        \epsfbox{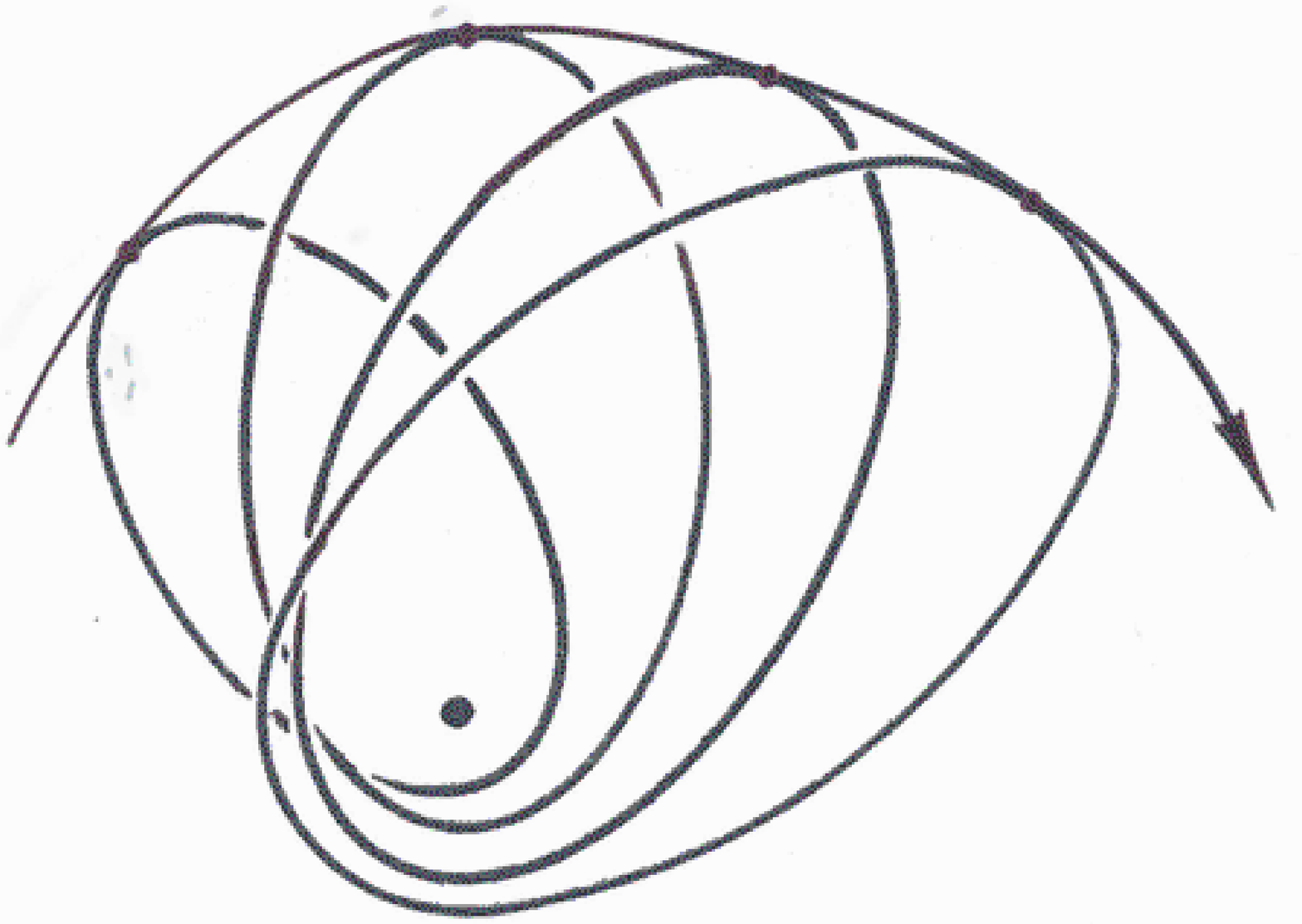}
 \end{center}
  \mbox{\small
 \parbox[b]{5.1in}{{\underline{Fig.3.}}~~~The orbit
 is represented by a sequence of confocal instantaneous ellipses
 that are
 tangent to the orbit, i.e., osculating. Now, the physical velocity
$\doterbold $ (tangent to the orbit) coincides with the
unperturbed Keplerian velocity $\bf\vec g$ (tangent to the
ellipse), so that their difference $\Phibold$ vanishes everywhere:
 \ba
 \nonumber
 \Phibold(t,\,C_1\,,\,.\,.\,.\,,\,C_6) \equiv\,
 \doterbold(t\,,\,C_1\,,\,.\,.\,.\,,\,C_6) -
 {\bf
  g}(t\,,\,C_1\,,\,.\,.\,.\,,\,C_6) =\,\sum_{j=1}^{6}\frac{\partial C_j}{\partial
 t}\,\dot{C}_j =0\,.
 \ea
 This equality, called Lagrange constraint or Lagrange gauge, is the necessary
 and sufficient condition of osculation.
}}

\subsection{Gauge-invariant planetary equations\\ of the Lagrange and Delaunay types}

We present the gauge-invariant Lagrange- and Delaunay-type
equations, following Efroimsky \& Goldreich (2003). These
equations follow from (\ref{30}) if we take into account the
gauge-invariance (i.e., the $\;\Phibold$-independence) of the
Lagrange-bracket matrix $\,[C_i\,C_j]\,$.
 \ba
 \nonumber
  \frac{da}{dt}\;=\;\frac{2}{n\,a}\;\;\left[\frac{\partial
 \left(\,-\,\Delta {\cal H} \right)}{\partial M_o} \;-\;\frac{\partial \,
 \Delta {\cal L}}{\partial \doterbold^{
 \left.~\right.} } \,
 \frac{\partial }{\partial M_o}
 \left({\Phibold}\,+\,\frac{
 \partial {\,\Delta {\cal L} }}{\partial \doterbold}\right)
 \;-\right.~~~~~~~~~~~~~~~~~~\\
 \label{46}
 \label{222}\\
 \nonumber
\left. \left({\Phibold}\,+\,\frac{\partial \,\Delta {\cal
L}}{\partial \doterbold}\right)\,\frac{\partial \bf {
g}}{\partial M_o}
  \;-\;\frac{\partial
 \efbold
  }{\partial
M_o}\, \;\frac{d}{dt} \left({\Phibold}\,+\,\frac{\partial \,\Delta
{\cal L}}{\partial \doterbold}\right) \right]\;\;\;,
 \ea
 \ba
 \nonumber\\
 \nonumber\\
 \nonumber
 \frac{de}{dt}\,=\,\frac{1-e^2}{n\,a^2\,e}\;\left[\frac{\partial
 \left(\,-\,\Delta {\cal H} \right)}{\partial M_o} \;-\;\frac{\partial \,\Delta
 {\cal L}}{\partial {\bf {\dot
{{\erbold}}}}^{
 \left.~\right.} } \,\frac{\partial }{\partial a}\left(\Phibold\,+\,
 \frac{\partial {\,\Delta {\cal L}}}{\partial
 \doterbold}\right) \,-\;~~~~~~~~~~~~~~~~~~~~~~~~~~~~~~\right.\\
 \nonumber\\
 \nonumber\\
 \nonumber
 \left.
 \left({\Phibold}\,+\,\frac{\partial \,\Delta {\cal L}}{\partial \doterbold}\right)\,
 \frac{\partial \bf {
 g}}{\partial M_o} \; -
\frac{\partial{\efbold}}{\partial M_o} \,\;\frac{d}{dt}
\left({\Phibold}\,+\,
 \frac{\partial {\,\Delta {\cal L}}}{\partial \doterbold}\right)
\right]\;-\;\;\;\;\;\;\;\;\;\;\;\;\;\;\;\;\;\;\;\\
 \label{47}
 \label{223}\\
 \nonumber\\
 \nonumber
\frac{(1\,-\,e^2)^{1/2}}{n\,a^2\,e} \;\left[\frac{\partial
 \left(\,-\,\Delta {\cal H} \right)}{\partial \omega} \;-\;
 \frac{\partial \,\Delta {\cal L}}{\partial \doterbold^{\left.~\right.} }
  \,\frac{\partial }{\partial \omega}\left({\Phibold}\,+
  \,\frac{\partial \,\Delta {\cal L}}{\partial \doterbold}\right)
   \;-~~~~~~~~\right.\\
 \nonumber\\
 \nonumber\\
 \nonumber
 \left.
 \left({\Phibold}\,+\,\frac{\partial \,\Delta {\cal L}}{\partial \doterbold}\right)\,
 \frac{\partial \bf {
 g}}{\partial
\omega}\;-\;\frac{\partial{\efbold}}{\partial \omega}
\,\;\frac{d}{dt} \left({\Phibold}\,+\,\frac{\partial \,\Delta
{\cal L}}{\partial \doterbold}\right) \right]\;\;\;\;,\;\;\;\;
 \ea
 ~\\
 \ba
 \nonumber
 \frac{d\omega}{dt}\;=\;\frac{\;-\;\cos \inc
}{n\,a^2\,(1\,-\,e^2)^{1/2}\, \sin \inc }\;\;\left[\frac{\partial
 \left(\,-\,\Delta {\cal H} \right)}{\partial \inc }
  \;-\;\frac{\partial \,\Delta {\cal L}}{\partial {\bf {\dot
{{\erbold}}}}^{
 \left.~\right.} }
  \,\frac{\partial }{\partial \inc}\left({\Phibold}\,
  +\,\frac{\partial \,\Delta {\cal L}}{\partial \doterbold}\right)
   \;-\;~~~~~~~~\right.\\
 \nonumber\\
 \nonumber\\
 \nonumber
 \left.
   \left({\Phibold}\,+\,\frac{\partial
 \,\Delta {\cal L}}{\partial \doterbold}\right)\,\frac{\partial \bf{
 g}}{\partial \inc } \;-\;\frac{\partial{\efbold}}{\partial
\inc }\, \;\frac{d}{dt} \left({\Phibold}\,+\,\frac{\partial
\,\Delta {\cal L}}{\partial \doterbold}\right)
 \right]\;+\;\;\;\\
   \label{48}
 \label{224}\\
 \nonumber\\
 \nonumber
 \frac{(1-e^2)^{1/2}}{n\,a^2\,e}\;
  \left[
  \frac{\partial  \left(\,-\,\Delta {\cal H} \right)}{\partial e} \;-\;
  \frac{\partial \,\Delta {\cal L}}{\partial {\bf {\dot {{\erbold}}}}^{
 \left.~\right.} } \,
 \frac{\partial }{\partial e}\left({\Phibold}\,+\,
 \frac{\partial \,\Delta {\cal L}}{\partial \doterbold}\right)
  \;-\;~~~~~~~~\right.\\
 \nonumber\\
 \nonumber\\
 \nonumber
 \left.
  \left({\Phibold}\,+\,\frac{\partial
\,\Delta {\cal L}}{\partial \doterbold}\right)\,\frac{\partial \bf
{
 g}}{\partial e} \;-\;\frac{\partial{\efbold}}{\partial e}\,
\;\frac{d}{dt} \left({\Phibold}\,+\,\frac{\partial \,\Delta {\cal
L}}{\partial \doterbold}\right)
 \right]\;\;\;,\;\;\;\;\;\;\;\;\;
 \ea
 ~\\
 \ba
 \nonumber
 \frac{d \inc }{dt}\;=\;\frac{\cos
\inc}{n\,a^2\,(1\,-\,e^2)^{1/2}\, \sin
\inc}\;\;\left[\frac{\partial  \left(\,-\,\Delta {\cal H}
\right)}{\partial \omega} \;-\;\frac{\partial \,\Delta {\cal
L}}{\partial {\bf {\dot {{\erbold}}}}^{
 \left.~\right.} } \,\frac{\partial }{\partial \omega}
 \left({\Phibold}\,+\,\frac{\partial \,\Delta {\cal L}}{\partial
 \doterbold}\right) \;-~~~~~~~~\right.\\
 \nonumber\\
 \nonumber\\
 \nonumber
 \left.
 \left({\Phibold}\,+\,\frac{\partial
\,\Delta {\cal L}}{\partial \doterbold}\right)\,\frac{\partial \bf
{
 g}}{\partial \omega } \;-\;\frac{\partial{\efbold}}{\partial
\omega }\, \;\frac{d}{dt} \left({\Phibold}\,+\,\frac{\partial
\,\Delta {\cal L}}{\partial
\doterbold}\right) \right]\;-\;\;\;\;\;\;\;\;\;\;\;\\
 \nonumber\\
   \label{49}
 \label{225}\\
 \nonumber
 \;\frac{1}{n\,a^2\,(1\,-\,e^2)^{1/2}\,\sin \inc
}\;\;\left[\frac{\partial  \left(\,-\,\Delta {\cal H}
\right)}{\partial \Omega} \;-\;\frac{\partial \,\Delta {\cal
L}}{\partial {\bf {\dot {{\erbold}}}}^{
 \left.~\right.} } \,\frac{\partial }{\partial \Omega}\left({\bf
  {
  \Phi}}\,+\,\frac{\partial \,\Delta {\cal L}}{\partial
   \doterbold}\right) \;-~~~~~~~~\right.\\
 \nonumber\\
 \nonumber\\
 \nonumber
 \left.
 \left({\Phibold}\,+\,\frac{\partial \,\Delta {\cal L}}{\partial \doterbold}\right)\,
 \frac{\partial \bf {
 g}}{\partial \Omega }
 \;-\;\frac{\partial \efbold}{\partial \Omega}\, \;\frac{d}{dt}
\left({\Phibold}\,+\,\frac{\partial
 \,\Delta {\cal L}}{\partial \doterbold}\right) \right]\;\;\;,\;\;
 \ea
 ~\\
 \ba
 \nonumber
\frac{d\Omega}{dt}\;=\;\frac{1}{n\,a^2\,(1\,-\,e^2)^{1/2}\,\sin
\inc }\;\; \left[\frac{\partial  \left(\,-\,\Delta {\cal H}
\right)}{\partial \inc } \;-\;\frac{\partial \,\Delta {\cal
L}}{\partial {\bf {\dot {{\erbold}}}}^{
 \left.~\right.} } \,\frac{\partial }{\partial \inc}\left({\bf
  {
  \Phi}}\,+\,\frac{\partial
   \,\Delta {\cal L}}{\partial \doterbold}\right) \;-~~~~~~~~\right.\\
 \nonumber\\
  \label{50}
 \label{226}\\
 \nonumber
 \left.\left( {\Phibold}\,+\,\frac{\partial \,\Delta {\cal L}}{\partial \doterbold}\right)\,
 \frac{\partial \bf {
 g}}{\partial \inc }
\;-\;\frac{\partial{\efbold}}{\partial \inc }\, \;\frac{d}{dt}
\left({\Phibold}\,+\,\frac{\partial \,\Delta {\cal L}}{\partial
\doterbold}\right)
 \right]\;\;\;,\;\;\;\;\;\;\;\;\;
 \ea
 ~\\
 \ba
 \nonumber
\frac{dM_o}{dt}\,=\,\;-\,\frac{1\,-\,e^2}{n\,a^2\,e}\,\;\left[
\frac{\partial  \left(\,-\,\Delta {\cal H} \right)}{\partial e }
\;-\;\frac{\partial \,\Delta {\cal L}}{\partial {\bf {\dot
{{\erbold}}}}^{ \left.~\right.} } \,\frac{\partial }{\partial
e}\left({\Phibold}\,+\,\frac{\partial \,\Delta {\cal L}}{\partial
\doterbold}\right) \,-~~~~~~~~\right.\\
 \nonumber\\
 \nonumber\\
 \nonumber
 \left.\left({\Phibold}\,+\,\frac{\partial \,\Delta {\cal L}}{\partial
 \doterbold}\right)\,\frac{\partial \bf {
g}}{\partial e} \,-\,\frac{\partial{\efbold}}{\partial e }
\,\;\frac{d}{dt} \left({\Phibold}\,+\,\frac{\partial \,\Delta
{\cal L}}{\partial \doterbold}\right)
 \right]\;-\\
  \label{51}
 \label{227}\\
 \nonumber\\
 \nonumber
 \frac{2}{n\,a}\,\left[\frac{\partial  \left(\,-\,\Delta {\cal H} \right)}{\partial a }
\;-\;\frac{\partial \,\Delta {\cal L}}{\partial {\bf {\dot
{{\erbold}}}}^{
 \left.~\right.} } \,\frac{\partial }{\partial a}
 \left({\Phibold}\,+\,\frac{\partial \,\Delta {\cal L}}{\partial
  \doterbold}\right) \,-~~~~~~~~\right.\\
 \nonumber\\
 \nonumber\\
 \nonumber
 \left. \left({\Phibold}\,+\,\frac{\partial \,\Delta {\cal L}}{\partial \doterbold}\right)\,
 \frac{\partial \bf {
  g}}{\partial a}
\,-\,\frac{\partial{\efbold}}{\partial a } \,\;\frac{d}{dt}
\left({\Phibold}\,+\,\frac{\partial \,\Delta {\cal L}}{\partial
\doterbold}\right)
 \right]\;\;\;\;.
 \ea

\pagebreak

Similarly, the gauge-invariant Delaunay-type system can be written
down as:
 \ba
\frac{dL}{dt}=\frac{\partial  \left(\,-\,\Delta {\cal
H}\right)}{\partial M_o}-\frac{\partial \,\Delta  L}{\partial
{\doterbold}} \,\frac{\partial }{\partial M_o}\left( \Phibold
\,+\,\frac{\partial \,\Delta {\cal L}}{\partial \doterbold}\right)
-\left(\Phibold\,+\,\frac{\partial \,\Delta {\cal L}}{\partial
\doterbold}\right)\, \frac{\partial {\bf
 g}}{\partial M_o}-\frac{\partial {\erbold}}{\partial
 M_o}\,\frac{d}{dt} \left(\Phibold\,+\,\frac{\partial \,\Delta
{\cal L}}{\partial \doterbold}\right)\;\;,\;\;
 \label{52}
 \label{Delaunay.1}\\
 \nonumber\\
\frac{d M_o}{dt}\;=\,-\,\frac{\partial \left(\,-\,\Delta {\cal
H}\right)}{\partial L}+\frac{\partial \,\Delta {\cal L}}{\partial
{\doterbold}} \,\frac{\partial }{\partial
L}\left({\Phibold}\,+\,\frac{\partial \,\Delta {\cal L}}{\partial
\doterbold}\right)+\left({{\bf\vec\Phi}}\,+\,\frac{\partial
\,\Delta {\cal L}}{\partial \doterbold}\right) \,\frac{\partial
{\bf
g}}{\partial L}+\frac{\partial {\erbold}}{\partial
L}\,\frac{d}{dt} \left({\Phibold}\,+\,\frac{\partial \,\Delta
{\cal L}}{\partial \doterbold}\right)\;\;\;\;,\;\;\;
 \label{53}
 \label{Delaunay.2}
 \ea
 \ba
 \frac{dG}{dt}\;=\;\frac{\partial \left(\,-\,\Delta {\cal H}\right)}{\partial
\omega}\;-\;\frac{\partial \,\Delta {\cal L}}{\partial
{\doterbold}} \,\frac{\partial }{\partial
\omega}\left({\Phibold}\,+\,\frac{\partial \,\Delta {\cal
L}}{\partial \doterbold}\right)\;-\;
 \left({{\Phibold}}\,+\,\frac{\partial \,\Delta {\cal L}}{\partial
\doterbold}\right)\;\frac{\partial {\bf
 g}}{\partial
\omega}\;-\;\frac{\partial {\erbold}}{\partial
\omega}\;\frac{d}{dt} \left({\Phibold}\,+\,\frac{\partial \,\Delta
{\cal L}}{\partial \doterbold}\right)\;\;\;,\;\;
  \label{54}
 \label{Delaunay.3}\\
 \nonumber\\
\frac{d\omega }{dt}\;=\,-\,\frac{\partial \left(\,-\,\Delta {\cal
H}\right)}{\partial G}\,+\,\frac{\partial \,\Delta {\cal
L}}{\partial {\doterbold}} \frac{\partial }{\partial
G}\left({\Phibold}\,+\,\frac{\partial \,\Delta {\cal L}}{\partial
\doterbold}\right)\,+\,\left({{\Phibold}}\,+\,\frac{\partial
\,\Delta {\cal L}}{\partial \doterbold}\right) \frac{\partial {\bf
 g} }{\partial G}\, +\,\frac{\partial {\erbold}}{\partial
G}\,\frac{d}{dt} \left({\Phibold}\,+\,\frac{\partial \,\Delta
{\cal L}}{\partial \doterbold}\right)\;\;\;,\;\;
  \label{55}
 \label{Delaunay.4}
 \ea
 \ba
 \frac{d H}{dt}\,=\,\frac{\partial \left(\,-\,\Delta {\cal H}\right)}{\partial
\Omega}\,-\,\frac{\partial \,\Delta {\cal L}}{\partial
{\doterbold}} \,\frac{\partial }{\partial
\Omega}\left({\Phibold}\,+\,\frac{\partial \,\Delta {\cal
L}}{\partial
\doterbold}\right)\,-\,\left({\Phibold}\,+\,\frac{\partial
\,\Delta {\cal L}}{\partial \doterbold}\right)\,\frac{\partial
{\bf
 g}}{\partial \Omega }\,-\,\frac{\partial \efbold}{\partial \Omega }\,\frac{d}{dt}
\left({\Phibold}\,+\,\frac{\partial \,\Delta {\cal L}}{\partial
\doterbold}\right)\;\;\;\;,\;\;\;
 \label{56}
  \label{Delaunay.5}\\
 \nonumber\\
\frac{d \Omega}{dt}\,=\,-\,\frac{\partial \left(\,-\,\Delta {\cal
H}\right)}{\partial H}\,+\,\frac{\partial \,\Delta {\cal
L}}{\partial {\doterbold}} \,\frac{\partial }{\partial
H}\left({\Phibold}\,+\,\frac{\partial \,\Delta {\cal L}}{\partial
\doterbold}\right)
 \,+\,\left({\Phibold}\,+\,\frac{\partial \,\Delta {\cal L}}{\partial
\doterbold}\right)\,\frac{\partial {\bf
 g}}{\partial
H}\,+\,\frac{\partial {\erbold}}{\partial H}\,\frac{d}{dt}
\left({\Phibold}\,+\,\frac{\partial \,\Delta {\cal L}}{\partial
\doterbold}\right)\;\;.\;\;
  \label{57}
 \label{Delaunay.6}
 \ea
where $\;\mu\;$ stands for the reduced mass, while
 \ba L\,\equiv\,\mu^{1/2}\,a^{1/2}\,\;\;,\;\;\;\;\;
G\,\equiv\,\mu^{1/2}\,a^{1/2}\,\left(1\,-\,e^2\right)^{1/2}\,\;\;,\;\;\;\;
H\,\equiv\,\mu^{1/2}\,a^{1/2}\,\left(1\,-\,e^2\right)^{1/2}\,\cos
\inc\,\;\;\;.\;\;\;
  \label{58}
 \label{229}
 \ea
The symbols $\;{\Phibold},\,{\efbold},\,{\bf{
  g}}\;$ now
denote the functional dependencies of the gauge, position, and
velocity upon the Delaunay, not Keplerian elements, and therefore
these are functions different from
$\;{\Phibold},\,{\efbold},\,{\bf{
 g}}\;$ used in (\ref{222} -
\ref{227}) where they stood for the dependencies upon the Kepler
elements. (In Efroimsky (2002a,b) the dependencies
$\;{\Phibold},\,{\efbold},\,{\bf{
 g}}\;$ upon the Delaunay
variables were equipped with tilde, to distinguish them from the
dependencies upon the Kepler coordinates.)

To employ the gauge-invariant equations in analytical calculations
is a delicate task: one should always keep in mind that, in case
$\,\Phibold\,$ is chosen to depend not only upon time but also
upon the ``constants" (but not upon their derivatives), the
right-hand sides of these equations will implicitly contain the
first derivatives $\,dC_i/dt\,$, and one will have to move them to
the left-hand sides (like in the transition from (\ref{22}) to
(\ref{23})). The choices $\;\Phibold\;=\;0\;$ and
$\,\Phibold\,=\;-\,{\partial \Delta{\cal L}}/{\partial
\doterbold}\,$ are exceptions. (The most general exceptional gauge
reads as $\;\Phibold\;=\;-\;{\partial \Delta{\cal L}}/{\partial
\doterbold}\;+\;\eta(t)\;$, where $\;\eta(t)\;$ is an arbitrary
function of time.)

As was expected from (\ref{30}), both the Lagrange and Delaunay
systems simplify in the gauge (\ref{31}). Since for orbital
motions we have $\;{\partial {\cal H}}/{\partial
{\pbold}}\;=\;-\;{\partial \Delta {\cal L}}/{\partial
\doterbold}\;$, then (\ref{31}) coincides with (\ref{44}). Hence,
the Hamiltonian analysis (\ref{34} - \ref{44}) explains why it is
exactly in the gauge (\ref{31}) that the Delaunay system becomes
symplectic. In physicists' parlance, the canonicity condition
breaks the gauge symmetry by stiffly fixing the gauge (\ref{44}),
gauge that is equivalent, in the orbital case, to (\ref{31}) --
phenomenon called ``gauge stiffness." The phenomenon may be looked
upon also from a different angle. Above we emphasised that the
gauge freedom implies essential arbitrariness in our choice of the
functional form of $\;\Phibold (t\,;\;C_1\,,\;.\;.\;.\;,\;C_6)\;$,
provided the choice does not come into a contradiction with the
equations of motion -- an important clause that shows its
relevance in (\ref{34} - \ref{44}) and (\ref{51} - \ref{56}): we
see that, for example, the Lagrange choice $\;\Phibold\,=\,0\;$
(as well as any other choice different from (\ref{31})) is
incompatible with the canonical structure of the equations of
motion for the elements.

\section{A practical example on gauges:\\ a satellite orbiting a precessing oblate planet.}

Above we presented the Lagrange- and Delaunay-type planetary
equations in the gauge-invariant form (i.e., for an arbitrary
choice of the gauge function
$\,\Phibold(t\,;\,C_1\,,\,.\,.\,.\,,\,C_6)\,$) and for a generic
perturbation $\,\Delta {\cal L}\,$ that may depend not only upon
positions but also upon velocities and the time. We saw that the
disturbing function is the negative Hamiltonian variation (which
differs from the Lagrangian variation when the perturbation
depends on velocities). Below, we shall also see that the
functional dependence of $\,\Delta {\cal H}\,$ upon the orbital
elements is gauge-dependent.

 \subsection{The Gauge Freedom and the Freedom of Frame Choice}

In the most compressed form, implementation of the
variation-of-constants method in orbital mechanics looks like
this. A generic solution to the two-body-problem is expressed with
 \begin{eqnarray}
 \erbold\;\;\;=\;\;\;{{\mbox{
 \boldmath$
 {f}$}}}\left(C,\,t \right)\;\;\;,\;\;\;\;\;
 \;\;\;\;\;\;\;\;\;\;\;\;\;\;
 \label{59}\\
 \nonumber\\
 \left(\frac{\partial{\mbox{\boldmath$
  {f}$}} }{\partial t}\right)_{C}=\;\;
 {\bf{
 {g}}}\left(C,\,t \right)
 \;\;\;,\;\;\;\;\;\;\;\;\;\;\;\;\;\;\;\;\;\;\;\;
 \label{60}\\
 \nonumber\\ \nonumber\\ \left(\frac{\partial {\bf{
 {g}}}}{\partial
 t}\right)_{C}=\;-\;\frac{\mu}{f^2}\;\frac{{\mbox{
 \boldmath$
 {f}$}}}{f}\;\;\;\;\;\;\;\;\;\;\;\;\;\;\;\;\;\;\;\;\;\;\;\;
 \label{61}
 \end{eqnarray}
and is used as an ansatz to describe the perturbed motion:
 \begin{eqnarray}
 {\erbold}\,&=&\,\efbold(C(t),\,t)\;\;,~~~~~~~~~~~~~~~~~~~~~~~~~~~~~~~~~~~~~~~~~
 \label{62}
 \ea
 \ba
 {\doterbold}\,&=&\,\frac{\partial{{{\f}}}}{\partial
 t}\;+\;\frac{\partial{{\f}}}{\partial C_i}\;\frac{d
 C_i}{dt}\;=\;{\bf{
 g}}\;+\;{\Phibold}  \;\;\;,~~~~~~~~~~~~~~~~~
 \label{63}
 \ea
 \ba
 {\bf{\ddot{
  r}}}\,= \,\frac{\partial {\bf{
  {g}}}}{\partial
 t}\;+\;\frac{\partial {\bf {
 g}}}{\partial C_i}\;\frac{d
 C_i}{dt}\;+\;\frac{d \Phibold }{dt}~
=\;-\;\frac{\mu}{f^2}\;\frac{{\mbox{
 \boldmath$
 {f}$}}}{f}\;+\;\frac{\partial {\bf
 g}}{\partial C_i}
 \;\frac{d
 C_i}{dt}\;+\;\frac{d \Phibold }{dt}\;\;\;.~~~~
 \label{64}
 \end{eqnarray}
As can be seen from (\ref{63}), our choice of a particular gauge
is equivalent to a particular way of decomposition of the physical
motion into a movement with velocity $\;{\bf{
g}}\;$ along the
instantaneous conic, and a movement caused by the conic's
deformation at the rate $\;\Phibold\,$.
 Beside the fact that we decouple the physical velocity $\;\doterbold\;$ in a certain
 proportion between these two movements, $\;{\bf{
 g}}\;$ and $\;\Phibold$, it also
 matters {\textbf{what}} physical velocity (i.e., velocity relative to what frame) is
 decoupled in this proportion. Thus, the choice of gauge does not exhaust all freedom: one
 can still choose {\textbf{in}} {\textbf{what}} {\textbf{frame}} to write
ansatz (\ref{62}) --  one can write it in inertial axes or in some
accelerated or/and rotating ones. For example, in the case of a
satellite orbiting a precessing oblate primary it is most
{\it{convenient}} to write the ansatz for the planet-related
position vector.

The kinematic formulae (\ref{62} - \ref{64}) do not yet contain
information about our choice of the reference system wherein to
employ variation of constants. This information shows up only when
(\ref{62}) and (\ref{64}) get inserted into the equation of motion
$\;{\bf{\ddot{\erbold}}}\,+\,(\mu\erbold/r^3)\,=\,{\Delta
\Fbold}\;$ to render
 \begin{equation}
 \frac{\partial {\bf {
 g}}}{\partial C_i}\;\frac{d
 C_i}{dt}\;+\;\frac{d \Phibold }{dt}\;=\;\Delta
{\efbold}\;=\;\frac{\partial \,\Delta {\cal L}}{\partial
 {\erbold}}\;-\;\frac{d}{dt}\,\left(\frac{\partial \,
 \Delta {\cal L}}{\partial
{\doterbold}}\right)\;\;\;.\;\;
 \label{65}
 \end{equation}
Information about the reference frame, where we employ the method
and define the elements $\,C_i\,$, is contained in the expression
for the perturbing force $\,\Delta {\efbold}$. If the operation is
carried out in an inertial system, $\,\Delta {\efbold}\,$ contains
only physical forces. If we work in a frame moving with a linear
acceleration $\,\bf\vec a\,$, then $\,\Delta {\efbold}\,$ also
contains the inertial force $\;-\,\bf\vec a\,$. In case this
coordinate frame also rotates relative to inertial ones at a rate
$\,\mubold\,$, then $\,\Delta {\efbold}\,$ also includes the
inertial contributions $\;-\,2\,{\mubold} \, \times \,
\doterbold\,-\,{\bf {\dot
{\mubold}}}\,\times\,\erbold\,-\,{\mubold}\times({\mubold}\times\doterbold)\,$.
When studying orbits about an oblate precessing planet, it is most
convenient (though not obligatory) to apply the
variation-of-parameters method in axes coprecessing with the
planet's equator of date: it is in this coordinate system that one
should write ansatz (\ref{62}) and decompose $\,\doterbold\,$ into
$\,{\bf{
g}}\,$ and $\,\Phibold\,$. This convenient choice of
coordinate system will still leave one with the freedom of gauge
nomination: in the said coordinate system, one will still have to
decide what function $\,\Phibold\,$ to insert in (\ref{63}).

\subsection{The disturbing function in a frame\\
co-precessing with the equator of date}

The equation of motion in the inertial frame is
\begin{equation}
 \erbold''\, =\;-\;
  \frac{\partial U}{\partial \erbold}\;\; ,
 \label{11.2}
 \label{66}
 \end{equation}
where U is the total gravitational potential, and time derivatives
in the inertial axes are denoted by primes. In a coordinate system
precessing at angular rate $\;\mubold(t)\;$, equation (\ref{66})
becomes:
 \begin{eqnarray}
 \nonumber
 {\bf{\ddot{
 {r}}}} \,=\,-\, \frac{\partial U}{\partial
 {\erbold}} \,-\, 2{\mubold} \, \times \,
 {\doterbold}\,-\,{\bf{\dot{\mubold}}}\,\times\,
 {\erbold}\,-\,{\mubold}\times({\mubold}\times
 {\erbold})~~~~~~~~~~~~~~~~~~~~
 \ea
 \ba
 =\;-\, \frac{\partial U_o}{\partial {\erbold}}
   \;-\;\frac{\partial \Delta U}{\partial {\erbold}}
 \,-\, 2{\mubold} \, \times \,
 {\doterbold}\,-\,{\bf{\dot{\mubold}}}\,\times\,
 {\erbold}\,-\,{\mubold}\times({\mubold}\times
 {\erbold})
 \;\;\;,
 \label{11.3}
 \label{67}
 \end{eqnarray}
 dots standing for time derivatives in the co-precessing frame,
 and $\;\mubold\;$ being the coordinate system's
 angular velocity relative to an inertial frame. Formula (\ref{A19}) in the Appendix
 gives the expression for $\;\mubold\;$ in terms of the longitude
 of the node and the inclination of the equator of date relative
 to that of epoch. The physical (i.e., not associated with inertial forces)
 potential $\,U(\erbold)\,$ consists of the (reduced) two-body part
 $\;U_o(\erbold)\,\equiv\,-\,G\,M\;\erbold/r^3
 \;$ and a term $\;\Delta U(\erbold)\;$ caused by the planet's oblateness (or, generally,
 by its triaxiality).

 To implement variation of the orbital
 elements defined in this frame, we note that the
 disturbing force on the right-hand side of (\ref{11.3}) is
 generated, according to (\ref{65}), by
 \begin{equation}
 \Delta {\cal L}\left({\erbold},\,{\doterbold},\,t
 \right)\,=\, -\;\Delta U(\erbold)\;+\;{\doterbold }{\bf \cdot}
 ({\mubold} \times {\erbold}) \;+\; \frac{1}{2}\;
 ({\mubold} \times {\erbold}){\bf \cdot}({\mubold}
 \times {\erbold})\;\;\;.
 \label{11.13}
 \label{68}
 \end{equation}
Since
 \begin{equation}
\frac{\partial \,\Delta {\cal L}}{\partial {\bf{\dot{r}}}}\;=\;
{\mubold}\times\erbold\;\;\;,
 \label{eq:dLddotr}
 \label{69}
 \end{equation}
 then
 \ba
 {\pbold}\;=\;\doterbold\;+\;\frac{\partial \,\Delta {\cal L}}{\partial
 {\bf{\dot{r}}}}\;=\;\doterbold\;+\;\mubold\,\times\,\erbold
 \label{70}
 \ea
and, therefore, the corresponding Hamiltonian perturbation reads:
 \begin{eqnarray}
 \nonumber
 \Delta {\cal H}\;=\;-\;\left[\Delta {\cal L}\;+\; \frac{1}{2}\left(\frac{\partial
 \,\Delta {\cal L}}{\partial {\bf{\dot{r}}}} \right)^2 \right]~~~~~~~~~~~~~~~~~~~~~~
~~~~~~~~~\\
   \label{eq:Hpert}
 \label{71}\\
 \nonumber
  =\;-\;\left[\,-\;\Delta U\;+\;{\pbold}\cdot ({\mubold}\times{\erbold})\;\right]\;=
 \;-\;\left[\,-\;\Delta U\;+\;({\erbold}\times{\pbold})\cdot\mubold\; \right]
 \;=\;\Delta U\;-\;{\bf
 J}\,\cdot\,\mubold\;\;\;,\\
 \nonumber
 \end{eqnarray}
 with vector $\;{\bf
 {J}}\;\equiv\;{\erbold}\times{\pbold}\;$
 being the satellite's orbital angular momentum in the inertial frame.

 According to (\ref{63}) and (\ref{70}), the momentum can be
 written as
 \begin{equation}
 {\pbold}\;=\;{\bf{
 {g}}}\;+\;\Phibold\;+\;{\mubold}\times{\efbold}\;\;\;,
 \label{72}
 \end{equation}
whence the Hamiltonian perturbation becomes
 \begin{eqnarray}
 \Delta {\cal H}\;=\;-\;\left[\Delta {\cal L}\;+\; \frac{1}{2}\left(\frac{\partial
 \,\Delta {\cal L}}{\partial {\doterbold}} \right)^2 \right]\;=\;-\;\left[\;-\;\Delta U
 \;+\;\left(
 {\efbold}\times {\bf{
 {g}}} \right)\cdot \mubold\;+\;
 \left( \Phibold \;+ \;{\mubold}\times{\efbold} \right) \cdot
 \left( {\mubold}\times{\efbold} \right)\;\right]\;\;.~~
 \label{73}
 \end{eqnarray}
This is what one is supposed to plug in (\ref{30}) or, the same,
in (\ref{46} - \ref{57}).

\subsection{Planetary equations in a precessing frame,\\
            written in terms of contact elements}

In the preceding subsection we fixed our choice of the frame
wherein to describe the orbit. By writing the Lagrangian and
Hamiltonian variations as (\ref{68}) and (\ref{73}), we stated
that our elements would be defined in the frame coprecessing with
the equator. The frame being fixed, we are still left with the
freedom of gauge choice. As evident from (\ref{33}) or (\ref{46} -
\ref{57}), the special gauge (\ref{31}) ideally simplifies the
planetary equations. Indeed, (\ref{31}) and (\ref{69}) together
yield
 \ba
 \Phibold\;=\;-\;\frac{\partial \,\Delta {\cal L}}{\partial
 \doterbold}\;=\;-\;{\mubold}\times\erbold\;\equiv\;-\;{\mubold}\times
 \efbold
 \;\;\;,
 \label{74}
 \ea
wherefrom the Hamiltonian (\ref{73}) becomes
 \begin{equation}
 \Delta {\cal H}^{(cont)}\;=\;-\;\left[\;-\;\Delta U(\efbold)+
 \mubold\cdot(\efbold\times {\bf {
 {g}}})\;\right]\;\;\;,
 \label{75}
 \end{equation}
while the planetary equations (\ref{30}) get the shape
 \begin{equation}
 [C_r\;C_i]\;\frac{dC_i}{dt}\;=\;
 \frac{\partial\;\left(\;-\;\Delta {\cal H}^{(cont)}\;\right)}{\partial C_r}\;\;\;\;,
 \label{76}
 \end{equation}
or, the same,
 \begin{equation}
 [C_r\;C_i]\;\frac{dC_i}{dt}\;=\;
 \frac{\partial}{\partial C_r}\;\left[\;-\;\Delta U(\efbold)\,+\,\mubold\cdot(
 \efbold\times {\bf {
 {g}}}) \;\right]\;\;\;\;,
 \label{77}
 \end{equation}
 where ${{\efbold}}$ and ${
 {\bf{g}}}$ stand for the undisturbed (two-body) functional
 expressions (\ref{59}) and (\ref{60}) of the position and velocity via the time and the chosen
 set of orbital elements. Planetary equations (\ref{76}) were obtained with aid of (\ref{74}),
 and therefore they render non-osculating orbital elements that are called contact elements. This
 is why we equipped the Hamiltonian (\ref{75}) with superscript ``{\it{(cont)}}." In distinction
 from the osculating elements, the contact ones osculate {\it{in phase space}}: (\ref{72}) and
 (\ref{74}) entail that $\;{\pbold}\;=\;{\bf
 g}\;$. As already mentioned in the end of
 section 1, existence of such elements was pointed out by Goldreich (1965) and Brumberg et al
 (1971) long before the concept of gauge freedom was introduced in celestial mechanics. Brumberg
 et al (1971) simply {\it{defined}} these elements by the condition that their insertion in
 $\;{\bf
 {g}}(t\,;\;C_1\,,\;.\;.\;.\;,\;C_6)\;$ returns not the perturbed velocity, but the
 perturbed momentum. Goldreich (1965) defined these elements (without calling them ``contact")
 differently. Having in mind inertial forces (\ref{67}), he wrote down the corresponding
 Hamiltonian (\ref{71}) and added its negative to the disturbing function of the standard
 planetary equations (without enriching the equations with any other terms). Then he noticed
 that those equations furnished non-osculating elements. Now we can easily see that both
 Goldreich's and Brumberg's definitions follow from the gauge choice (\ref{31}).

 When one chooses the Keplerian parameterisation, then (\ref{77}) becomes:
  \begin{equation}
\frac{da}{dt}\;=\;\frac{2}{n\,a}\;\;\frac{\partial\left(\,-\,\Delta
{\cal H}^{(cont)}\right) }{\partial
M_o}\;\;\;\;\;,\;\;\;\;\;\;\;\;~~~~~~~~~~~~~~~~~~~~~~~~~~~~~~~~~~~~~~~~~~~~~~~~~~
 ~~~~~~~~~~~~~~~~~~~~~~~~~~~~~~~
 \label{78}
 \label{446}
  \end{equation}
 ~\\
  \begin{eqnarray}
 \frac{de}{dt}\,=\,\frac{1-e^2}{n\,a^2\,e}\;\;\frac{\partial
 \left(\,-\,\Delta {\cal H}^{(cont)} \right)  }{\partial M_o}\;-\;
\frac{(1\,-\,e^2)^{1/2}}{n\,a^2\,e} \;\frac{\partial
\left(\,-\,\Delta {\cal H}^{(cont)} \right) }{\partial \omega}
\;\;\;,\;~~~~~~~~~~~~~~~~~~~~~~~~~~~~~~~
 \label{79}
 \label{447}
  \end{eqnarray}
 ~\\
  \begin{eqnarray}
 \frac{d\omega}{dt}\,=\,\frac{\;-\,\cos \inc
}{n\,a^2\,(1\,-\,e^2)^{1/2}\, \sin \inc }\;\frac{\partial
\left(\,-\,\Delta {\cal H}^{(cont)} \right) }{\partial \inc }
\,+\,\frac{(1-e^2)^{1/2}}{n\,a^2\,e}\;\frac{\partial
\left(\,-\,\Delta {\cal H}^{(cont)} \right) }{\partial e}
~~~~~~~~~~~~~~~~~~~
 \label{80}
 \label{448}
  \end{eqnarray}
 ~\\
  \begin{eqnarray}
 \frac{d \inc }{dt}\,=\,\frac{\cos
\inc}{n\,a^2\,(1\,-\,e^2)^{1/2}\, \sin \inc}\;\;\frac{\partial
\left(\,-\,\Delta {\cal H}^{(cont)} \right) }{\partial \omega} \;-
 \;\frac{1}{n\,a^2\,(1\,-\,e^2)^{1/2}\,\sin \inc
}\;\;\frac{\partial \left(\,-\,\Delta {\cal H}^{(cont)} \right)
}{\partial \Omega} \;\;\,,\,\;~
 \label{81}
 \label{449}
  \end{eqnarray}
 ~\\
 ~\\
  \begin{eqnarray}
 \frac{d\Omega}{dt}\;=\;\frac{1}{n\,a^2\,(1\,-\,e^2)^{1/2}\,\sin
 \inc }\;\;\frac{\partial \left(\,-\,\Delta {\cal H}^{(cont)}
 \right) }{\partial \inc }
 \;\;\;,\;\;\;\;\;\;~~~~~~~~~~~~~~~~~~~~~~~~~~~~~~~~~~~~~~~~~~~~~~~~
 \label{82}
 \label{450}
  \end{eqnarray}
 ~\\
 ~\\
  \begin{eqnarray}
\frac{dM_o}{dt}\,=\,\;-\,\frac{1\,-\,e^2}{n\,a^2\,e}\,\;
\frac{\partial \left(\,-\,\Delta {\cal H}^{(cont)} \right)
 }{\partial e } \;-\;
 \frac{2}{n\,a}\,\frac{\partial \left(\,-\,\Delta {\cal H}^{(cont)} \right)
 }{\partial a }
\;\;\;\;.\;\;~~~~~~~~~~~~~~~~~~~~~~~~~~~~~~~~~
 \label{83}
 \label{451}
  \end{eqnarray}
The above equations implement an interesting pitfall. When
describing orbital motion relative to a frame coprecessing with
the equator of date, it is tempting to derive the Hamiltonian
variation caused by the inertial forces, and to simply plug it,
with a negative sign, into the disturbing function. This would
entail equations (\ref{76} - \ref{83}) which, as demonstrated
above, belong to the non-Lagrange gauge (\ref{31}). The elements
furnished by these equations are nonosculating, so that the conics
parameterised by these elements are not tangent to the perturbed
trajectory. For example, $\,\inc\,$ gives the inclination of the
instantaneous non-tangent conic, but differs from the real,
physical physical (i.e., osculating), inclination of the orbit.
This approach -- when an inertial term is simply added to the
disturbing function -- was employed by Goldreich (1965), Brumberg
et al (1971), and Kinoshita (1993), and many others. Goldreich and
Brumberg noticed that this destroyed osculation, Kinoshita failed
to.

Goldreich (1965) studied how the equinoctial precession of Mars
influences the long-term evolution of Phobos' and Deimos' orbit
inclinations. Goldreich assumed that the elements $\;a\;$ and
$\;e\;$ stay constant; he also substituted the Hamiltonian
variation (\ref{75}) with its orbital average, which made his
planetary equations render the secular parts of the elements. He
assumed that the averaged physical term $\;\bra\,\Delta U\,\ket\;$
is only due to the primary's oblateness:
 \ba
 \bra\,\Delta U\,\ket\;=\;-\;\frac{n^2\;J_2}{4}\;\,{\rho}^2\;\;
 \frac{3\;\cos^2i\;-\;1}{\left(1\;-\;e^2\right)^{3/2}}\;\;\;,
 \label{84}
 \ea
$\rho\;$ being the mean equatorial radius of the
planet,\footnote{~Goldreich used the nonsphericity parameter
$\,J\equiv\left({\textstyle 3}/{\textstyle
2}\right)\,\left({\textstyle \rho_e}/{\textstyle \rho}\right)^2
J_2\,$, where $\,\rho_e\,$ is the mean {\it{equatorial}} radius.}
and $\;n\;$ being the satellite's mean motion. To simplify the
inertial term, Goldreich employed the well known formula
 \ba
  \erbold\,\times\,{\bf
  g}\;=\; \sqrt{G\,m\;a\;\left(1\;-\;e^2\right)}\;\,{\wbold}\;\;\;,
  \label{85}
  \ea
where
 \ba
 {\wbold}\;=\;{\bf\hat{x}}_1\;\sin \inc\;\sin \Omega\;-
 \;{\bf\hat{x}}_2\;\sin
 \inc\;\cos \Omega\;+\;{\bf\hat{x}}_3\;\cos
 \inc\;\;~~~~~~~~~~~~~~~~~~~
 \label{86}
 \ea
is a unit vector normal to the instantaneous ellipse, expressed
through unit vectors
$\;{\bf\hat{x}}_1,\,{\bf\hat{x}}_2,\,{\bf\hat{x}}_3\;$ associated
with the co-precessing frame $\,\it x_1,\,x_2,\,x_3\,$ (axes
$\,\it x_1\,$ and $\,\it x_2\,$ lying in the planet's equatorial
plane of date, and $\,\it x_1\,$ pointing along the fiducial line
wherefrom the longitude of the ascending node of the satellite
orbit, $\;\Omega\;$, is measured). This resulted in
 \begin{eqnarray}
 \nonumber
 \bra\,\Delta {\cal H}^{(cont)}\,\ket\;=\;-\;\left[\;-\;\bra \Delta U \ket\;+\;
 \bra\, \mubold\cdot(\efbold\times {\bf {
 {g}}})\,\ket \;\right]\;=
 ~~~~~~~~~~~~~~~~~~~~~~~~~~~~\\
 \label{87}\\
 \nonumber
 -\;\frac{G\,m\,J_2}{4}\;\;\frac{\rho_e^2}{a^3}\;\;
 \frac{3\,\cos^2i\,-\,1}{\left(1\,-\,e^2\right)^{3/2}}\,-\,
 \sqrt{G\,m\,a\,\left(1\,-\,e^2\right)}\;\;\left(\,
 \mu_1  \;\sin i\;\sin \Omega\,-\, \mu_2  \;\sin i\;\cos \Omega
 \,+\, \mu_3 \;\cos i \,\right)\,~,~~
 \end{eqnarray}
all letters now standing not for the appropriate variables but for
their orbital averages. Substitution of this averaged Hamiltonian
in (\ref{81} - \ref{82}) lead Goldreich, in assumption that both
$\;|\dotmubold|/\left(\,n^2\,J_2\,\sin\inc\,\right)\;$ and
$\;|\mubold|/\left(\,n\,J_2\,\sin\inc\, \right)\;$ are much less
than unity, to the following system:
 \begin{eqnarray}
 \frac{d\Omega}{dt}\,\approx~
 -\;\frac{3}{2}\,{n\,J_2}\,\left(\frac{\rho_e}{a}\right)^2
 \;\frac{\cos
 {i}}{\left(1-e^2
 \right)^2}\;\;\;,
 \label{88}
 \end{eqnarray}
  ~\\
 \begin{eqnarray}
 \frac{d \inc}{dt}\;\approx\;-\;\mu_1\;\cos \Omega\;-\;
 \mu_2\;\sin \Omega\;\;\;{,}~~~
 \label{89}
 \end{eqnarray}
whose solution,
 \ba
 \nonumber
 i\,=\;-\,\frac{\mu_1}{\chi}\,\cos \left[\;-\,\chi\,
 \left(t\,-\,t_o\right)\,+\,\Omega_o
 \right]\,+\,\frac{\mu_2}{\chi}\,\sin \left[\;-\,\chi\,
 \left(t\,-\,t_o\right)\,+\,\Omega_o
 \right]\,+\,i_o \;\;\;{,}\\
 \label{90}\\
 \nonumber
 \Omega\,=\,-\,\chi\,\left(t\,-\,t_o\right)\,+\,\Omega_o~~~~~~~
 \mbox{where}\;\;\;~~\chi\;\equiv\;
 \frac{3}{2}\;{n\;J_2}\;\left(\frac{\rho_e}{a}\right)^2\;\;
 \frac{\cos {\inc}}{\left(1\;-\;e^2
 \right)^2}\;\;\;,~
 \ea
 tells us that in the course of equinoctial precession the satellite
 inclination oscillates about $\inc_o\,$.

 Goldreich (1965) noticed that his $\;\inc\;$ and the other elements were not
 osculating, but he assumed that their secular parts would differ from those of the
 osculating ones only in the orders higher than $\,O(|\mubold|)\,$. Below we shall probe
 the applicability limits for this assumption. (See the end of subsection 3.5.)

\subsection{Planetary equations in a precessing frame,\\ in terms of osculating elements}

When one introduces elements in the precessing frame and also
demands that they osculate in this frame (i.e., obey the Lagrange
constraint $\,\Phibold\,=\,0\,$), the Hamiltonian variation
reads:\footnote{~Both $\;\Delta {\cal H}^{(cont)}\;$ and $\;\Delta
{\cal H}^{(osc)}\;$ are equal to $\;\;-\,\left[\;-\;\Delta
U({{\efbold}},\,t)\,+\,
 \mubold\cdot{\bf{
 J}}\right]\,=\,-\,\left[\;-\;\Delta U({{\efbold}},\,t)\,+\,
 \mubold\cdot(\efbold\times{\pbold })\right]\;$. However, the
 canonical momentum now is different from $\;\bf
 g\;$ and reads as:
 $\;\pbold\;=\;{\bf{
 g}}\;+\;\left(\mubold\times\efbold\right)\;$. Hence, the
 functional forms of $\;\Delta {\cal H}^{(osc)}(\efbold\,,\;\pbold)\;$ and
 $\;\Delta {\cal H}^{(can)}(\efbold\,,\;\pbold)\;$ are different, though their values
 coincide.}
 \begin{equation}
 \Delta {\cal H}^{(osc)}\;=\;-\,\left[\;-\;\Delta U\,+\,
 \mubold\cdot(\efbold\times{\bf{
 g}})\;+\;
 (\mubold\times\efbold)\cdot(\mubold\times\efbold)\;\right]\;\;\;,
 \label{eq:DelHosc}
 \label{91}
 \end{equation}
 while equation (\ref{30}) becomes:
 \begin{eqnarray}
 \nonumber [C_n\;C_i]\;\frac{dC_i}{dt}\;=\;-\;
 \frac{\partial\,\Delta {\cal H}^{(osc)}}{\partial
 C_n}\,~~~~~~~~~~~~~~~~~~~~~~~~~
 \nonumber \\ &&
 \label{92}\\
 \nonumber +\;\mubold\cdot \left(\frac{\partial{\efbold}}{\partial
 C_n}\times {\bf{
 g}}\;-\;{\efbold}\times \frac{\partial{
 {\bf
 g}}}{\partial C_n}\right)\;-\; {\bf{\dot{\mubold
 }}}\cdot\left(\efbold\times \frac{\partial \efbold }{\partial
 C_n}\right)\;-\;\left(\mubold\times\efbold\right) \;\frac{\partial
 }{\partial C_n}\left(\mubold\times\efbold\right) \;\;\;.
 \end{eqnarray}
To ease the comparison of this equation with (\ref{77}), it is
convenient to split the expression (\ref{91}) for $\;\Delta {\cal
H}^{(osc)}\;$ into two parts:
 \ba
\Delta {\cal H}^{(cont)}\;=\;-\,\left[\;R_{oblate}({{\efbold
}},\,t)\;+\;\mubold\cdot(\efbold \times {\bf{
g}})\;\right]
 \label{93}
 \ea
 and
 \ba
 -\,(\mubold\times\efbold)\cdot(\mubold\times\efbold)\;\;\;,
 \label{94}
 \ea
 and then to group the latter part with the last term on the right-hand side of (\ref{35}):
 \begin{eqnarray}
 \nonumber [C_n\;C_i]\;\frac{dC_i}{dt}\;=\;-\;
 \frac{\partial\;\Delta {\cal H}^{(cont)}}{\partial C_n}~~~~~~~~~~~~~~~~
 \nonumber \\ &&
 \label{eq:avoposc}
 \label{95}\\
 \nonumber
 +\;\mubold\cdot
 \left(\frac{\partial{\efbold}}{\partial C_n}\times
 {\bf{
 g}}\;-\;{\efbold}\times
 \frac{\partial{
 {\bf g}}}{\partial C_n}\right)\;-\;
 {\bf{\dot{\mubold }}}\cdot\left(\efbold\times
 \frac{\partial \efbold }{\partial C_n}\right)\;+\;\left(\mubold\times
\efbold\right)
 \;\frac{\partial }{\partial C_n}\left(\mubold\times\efbold\right) \;\;\;.
 \end{eqnarray}
 Comparison of this analytical theory with a straightforward numerical
 integration\footnote{~Credit for this comparison goes to Pini Gurfil and Valery Lainey.}
 has confirmed that the $\;O(|\mubold|^2)\;$ term in (\ref{95}) may be
 neglected over time scales of, at least, hundreds of millions of years. In this
 approximation there is no the difference between $\;\Delta {\cal H}^{(cont)}\;$ and
 $\;\Delta {\cal H}^{(osc)}\;$, so we shall write down the equations as:
 \begin{eqnarray}
 [C_n\;C_i]\;\frac{dC_i}{dt}\;=\;-\;
 \frac{\partial\;\Delta {\cal H}^{(cont)}}{\partial C_n}~
 +\;\mubold\cdot
 \left(\frac{\partial{\efbold}}{\partial C_n}\times
 {\bf{
 g}}\;-\;{\efbold}\times
 \frac{\partial{\vec{\bf g}}}{\partial C_n}\right)\;-\;
 {\bf{\dot{\mubold }}}\cdot\left(\efbold\times
 \frac{\partial \efbold }{\partial C_n}\right)\;\;\;.\;\;\;\;
 \label{96}
 \end{eqnarray}
For $\;{C_i}\;$ chosen as the Kepler elements, inversion of the
Lagrange brackets in (\ref{90}) will yield the following
Lagrange-type system:
~\\
 \ba
 \frac{da}{dt}\,=\,
 \frac{2}{n a}\,\left[\,\;\frac{\partial\left(\,-\,\Delta {\cal H}^{(cont)}\right)
 }{\partial M_o}\,-\,
 {\bf{\dot{\mubold }}}\cdot\left(\efbold\times
 \frac{\partial \efbold }{\partial
 M_o}\right)\;\right]\;\;,\;\;\;~~~~~~~~~~~~~~~~~~~~~~~~~~~~~~~~~~~~~~~~~~~
 \label{97}
 \label{459}
  \ea
  ~\\
 ~\\
  \begin{eqnarray}
  \nonumber
 \frac{de}{dt}\,=\,\frac{1-e^2}{n\,a^2\,e}\;\;\left[\;\frac{\partial
 \left(\,-\,\Delta {\cal H}^{(cont)} \right)  }{\partial M_o}
 \,-\,
 {\bf{\dot{\mubold }}}\cdot\left(\efbold\times
 \frac{\partial \efbold }{\partial M_o}\right)
 \;\right]~~\,~~~~~~~~~~~~~~~~~~~~~~~~~~~~~~~~~~~~~~~~~\\
 \nonumber\\
 \label{460}
 \label{98}\\
 \nonumber\\
 \nonumber
 -\;
 \frac{(1\,-\,e^2)^{1/2}}{n\,a^2\,e} \;\left[\;\frac{\partial
 \left(\,-\,\Delta {\cal H}^{(cont)} \right) }{\partial \omega} \,~ +\, \mubold\cdot
 \left(\frac{\partial{\efbold}}{\partial \omega}\times
 {\bf{
 g}}\,-\,{\efbold}\times
 \frac{\partial{
 {\bf g}}}{\partial \omega}\right)\,-\,
 {\bf{\dot{\mubold }}}\cdot\left(\efbold\times
 \frac{\partial \efbold }{\partial \omega}\right)
 \;\right]~~,~~~
  \end{eqnarray}
 ~\\
 ~\\
  \begin{eqnarray}
 \nonumber
 \frac{d\omega}{dt}\,=~
 \frac{\;-\,\cos \inc
}{n a^2(1-e^2)^{1/2}\, \sin \inc }\left[ \frac{\partial
\left(\,-\,\Delta {\cal H}^{(cont)} \right) }{\partial \inc } \,~
+ ~ \mubold\cdot
 \left(\frac{\partial{\efbold}}{\partial \inc}\times
 {\bf{
 g}}\,-\,{\efbold}\times
 \frac{\partial{
 {\bf g}}}{\partial \inc}\right)\,-\,
 {\bf{\dot{\mubold }}}\cdot\left(\efbold\times
 \frac{\partial \efbold }{\partial \inc}\right)\,
  \right]~~~~\\
 \nonumber\\
 \label{99}\\
 \nonumber\\
 \nonumber
 +\,~\frac{(1-e^2)^{1/2}}{n\,a^2\,e}\;
 \left[\;\frac{\partial \left(\,-\,\Delta {\cal H}^{(cont)} \right) }{\partial e}\,
+ ~ \mubold\cdot
 \left(\frac{\partial{\efbold}}{\partial e}\times
 {\bf{
 g}}\,-\,{\efbold}\times
 \frac{\partial{
 {\bf g}}}{\partial e}\right)\,-\,
 {\bf{\dot{\mubold }}}\cdot\left(\efbold\times
 \frac{\partial \efbold }{\partial e}\right)
 \;\right]\;\;,~~~~~~~~~~~
  \end{eqnarray}
 \ba
 \nonumber
 \frac{d \inc
 }{dt}\;=~ \frac{\cos \inc}{n a^2\,(1 - e^2)^{1/2} \sin
 \inc}
 \left[\frac{\partial \left(\,-\,\Delta {\cal H}^{(cont)} \right)}{\partial
 \omega}\;+ \; \mubold\cdot
 \left(\frac{\partial{\efbold}}{\partial \omega}\times
 {\bf{
 g}}\,-\,{\efbold}\times
 \frac{\partial{
 {\bf g}}}{\partial \omega}\right)\,-\,
 {\bf{\dot{\mubold }}}\cdot\left(\efbold\times
 \frac{\partial \efbold }{\partial \omega}\right)
 \,\right]\,-\;\;\;\\
 \nonumber\\
 \label{100}\\
 \nonumber\\
 \nonumber
 \frac{1}{na^2\,(1-e^2)^{1/2}\,\sin \inc
 }\,\left[\,\frac{\partial \left(\,-\,\Delta {\cal H}^{(cont)} \right) }{\partial
 \Omega}\;+ \; \mubold\cdot
 \left(\frac{\partial{\efbold}}{\partial \Omega}\times
 {\bf{
 g}}\,-\,{\efbold}\times
 \frac{\partial{
 {\bf g}}}{\partial \Omega}\right)\,-\,
 {\bf{\dot{\mubold }}}\cdot\left(\efbold\times
 \frac{\partial \efbold }{\partial \Omega}\right)
 \,\right]\;{,}\;\;\;
  \ea
 ~\\
 ~\\
  \ba
\frac{d\Omega}{dt}\,=~\frac{1}{n a^2\,(1-e^2)^{1/2}\,\sin \inc
}\,\left[\,\frac{\partial \left(\,-\,\Delta {\cal H}^{(cont)}
\,\right) }{\partial \inc }\,~ + ~\mubold\cdot
 \left(\frac{\partial{\efbold}}{\partial \inc}\times
 {\bf{
 g}}\,-\,{\efbold}\times
 \frac{\partial{
 {\bf g}}}{\partial \inc}\right)\,-\,
 {\bf{\dot{\mubold }}}\cdot\left(\efbold\times
 \frac{\partial \efbold }{\partial \inc}\right)\,
 \right]\;,\;\;\;
 \label{101}
  \ea
 ~\\
 ~\\
  \begin{eqnarray}
  \nonumber
\frac{dM_o}{dt}\,=\,\;-\,\;\frac{1\,-\,e^2}{n\,a^2\,e}\,\;
\left[\;\frac{\partial \left(\,-\,\Delta {\cal H}^{(cont)} \right)
 }{\partial e }~ + ~
 \mubold\cdot
 \left(\frac{\partial{\efbold}}{\partial e}\times
 {\bf{
 g}}\,-\,{\efbold}\times
 \frac{\partial{
 {\bf g}}}{\partial e}\right)\,-\,
 {\bf{\dot{\mubold }}}\cdot\left(\efbold\times
 \frac{\partial \efbold }{\partial e}\right)
 \;\right] ~~~~~\\
 \nonumber\\
 \label{102}\\
 \nonumber\\
 \nonumber
 -\;~\frac{2}{n\,a}\,\left[\;\frac{\partial \left(\,-\,\Delta {\cal
 H}^{(cont)}
 \right)
 }{\partial a }~ + ~ \mubold\cdot
 \left(\frac{\partial{\efbold}}{\partial a}\times
 {\bf{
 }}\,-\,{\efbold}\times
 \frac{\partial{
 {\bf g}}}{\partial a}\right)\,-\,
 {\bf{\dot{\mubold }}}\cdot\left(\efbold\times
 \frac{\partial \efbold }{\partial a}\right)
 \;\right]
 \;\;,~~~~~
  \end{eqnarray}
  ~\\
  terms $\;\mubold\cdot\left(\;(\partial \efbold/\partial M_o)\times {\bf
  g}\,-\,
  (\partial {\bf
  g}/\partial M_o)\times\efbold\;\right)\;$ being
  omitted in (\ref{97} - \ref{98}), because these terms vanish
  identically (see the Appendix to Efroimsky (2004)$\,$).

 \subsection{Comparison of calculations performed in the two above gauges}

Simply from looking at (\ref{76} - \ref{83}) and (\ref{96} -
\ref{102}) we notice that the difference in orbit descriptions
performed in the two gauges emerges already in the first order of
the precession rate $\;\mubold\;$ and in the first order of
$\;\dotmubold\;$.

Calculation of the $\;\mubold$- and $\;\dotmubold$-dependent terms
emerging in (\ref{97} - \ref{102}) takes more than 20 pages of
algebra. The resulting expressions are published in Efroimsky
(2005a), their detailed derivation being available in web-archive
preprint Efroimsky (2004). As an illustration, we present a couple
of expressions:
  \ba
 \nonumber
-\;\dotmubold\;\cdot\;\left(\;\efbold\;\times\;\frac{\partial
\efbold}{\partial
 \inc} \;\right)\;=\;
 ~~~~~~~~~~~~~~~~~~~~~~~~~~~~~~~~~~~~~~~~~~~~~~~~~~~~~~~~~~~~~~
 ~~~~~~~~~~~~~~~~~~\\
 \nonumber\\
 \nonumber\\
 \nonumber
 a^2\;\;
 \frac{\left(1\;-\;e^2\right)^2}{\left(1\;+\;e\;\cos
\nu\right)^2}\;\;\left\{\;\;\dot{\mu}_{1}\;
 \;\left[\;
           -\;\cos\Omega\;\sin(\omega +\nu)\;-\;
           \sin \Omega\;\cos(\omega +\nu)\;\cos \inc
  \;\right]\;\sin(\omega + \nu)\;  \right.\\
  \nonumber\\
  \nonumber\\
  \nonumber
+\;\dot{\mu}_2\;
  \left[\;-\;\sin\Omega\;\sin(\omega +\nu)\;+\;
                \cos \Omega\;\cos(\omega +\nu)\;\cos \inc
                   \;\right]\;\sin (\omega + \nu)\;~~\\
 \nonumber\\
 \nonumber\\
 \left.
+\;\dot{\mu}_3\;\sin(\omega + \nu)\;\cos(\omega + \nu)\;\sin \inc
 \;\;\right\}~~~,~~~~~~~~~~~~~~~~~~~~~~~~~~~~~~~~~~~~~
 \label{103}
 \ea
 ~\\
  \ba
 \mubold\;\cdot\;\left(\; \frac{\partial \efbold}{\partial e}\;\times\;{\bf
g}\;-\;\efbold\;\times\;\frac{\partial {\bf \vec g}}{\partial
e}\;\right)\;=\;
 -\;\,\mu_{\perp}\;\,\frac{n\,a^2\;\left(3\,e\,+\,2\,\cos \nu\,+\,e^2\,\cos \nu
 \right)}{\left(1\,+\,e\,\cos
 \nu\right)\;\sqrt{1\;-\;e^2}}
 ~~~,~~~~~~~~
 \label{104}
 \label{476}
 \ea
$\nu\;$ denoting the true anomaly. The fact that almost none of
these terms vanish reveals that equations (\ref{76} - \ref{83})
and (\ref{96} - \ref{102}) may yield very different results, i.e.,
that the contact elements may differ from their osculating
counterparts already in the first order of $\;\mubold\;$.

Luckily, in the practical situations we need not the elements
{\it{per se}} but their secular parts. To calculate these, one can
substitute both the Hamiltonian variation and the $\;\mubold$- and
$\;\dotmubold$-dependent terms with their orbital
averages\footnote{~Mathematically, this procedure is, to say the
least, not rigorous. In practical calculations it works well, at
least over not too long time scales.} calculated through
 \ba
 \langle\;\,.\,.\,.\,\;\rangle\;\equiv\;\frac{\left(1\;-\;e^2\right)^{3/2}}{2\;\pi}\;
 \int_{0}^{2\pi}\;.\,.\,.\;\;\;\frac{d\nu}{\left(1\;+\;e\;\cos \nu
 \right)^2}\;\;\;.
 \label{105}
 \ea
The situation might simplify very considerably if we could also
assume that the precession rate $\;\mubold\;$ stays constant. Then
in equations (\ref{97} - \ref{102}), we would assume
$\;\mubold\;=\;const\;$ and proceed with averaging the expressions
$\;\left(\;(\partial \efbold/\partial C_j)\times {\bf
g}\,-\,\efbold\times(\partial {\bf
g}/\partial
C_j)\;\right)\;$ only (while all the terms with $\;\dotmubold\;$
will now vanish).

Averaging of the said terms is lengthy and is presented in the
Appendix to Efroimsky (2004). All in all, we get, for constant
$\;\mubold\;$:
  ~\\
 \ba
 \mubold\,\cdot\,\langle\;\left(\;\frac{\partial \efbold}{\partial a}\,\times\,{\bf \vec
g}\,-\,\efbold\,\times\,\frac{\partial {\bf
g}}{\partial
a}\;\right)\;\rangle\,=\,\mubold\,\cdot\,\left(\;\frac{\partial
\efbold}{\partial a}\,\times\,{\bf \vec
g}\,-\,\efbold\,\times\,\frac{\partial {\bf
g}}{\partial
a}\;\right) \,=\,\frac{3}{2}\;\;
\mu_{\perp}\;\;\sqrt{\frac{G\;m\;\left(1\;-\;e^2\right)}{a}}\;\;\;,~~
 \label{106}
 \label{487}
 \ea
~\\
  \ba
 \mubold\;\cdot\;\langle\;\left(\; \frac{\partial \efbold}{\partial C_j}\;\times\;{\bf
 g}\;-\;\efbold\;\times\;\frac{\partial {\bf
 g}}{\partial
C_j}\;\right)\;\rangle\;=\;0\;\;\;~,~~~~~~~~~C_j\;=\;e\,,\;\Omega\,,\;\omega\,,\;\inc\,,
\;M_o\;\;\;. ~~~~~~~~~~~~~~~~~~~~
 \label{107}
 \label{488}
 \ea
Since the orbital averages (\ref{107}) vanish, then $\;e\;$ will,
along with $\;a\;$, stay constant for as long as our approximation
remains valid. Besides, no trace of $\;\mubold\;$ will be left in
the equations for $\;\Omega\;$ and $\;\inc\;$. This means that, in
the assumed approximation and under the extra assumption of
constant $\;\mubold\;$, the afore quoted analysis (\ref{84} -
\ref{90}), offered by Goldreich (1965), will remain valid at time
scales which are not too long.

 In the realistic case of time-dependent
precession, the averages of terms containing $\;\mubold\;$ and
$\;\dotmubold\;$ do not vanish (except for
$\;\mubold\cdot\left(\;(\partial \efbold/\partial M_o)\times {\bf
g}\,-\,\efbold\times(\partial {\bf
g}/\partial
M_o)\;\right)\;$, which is identically nil). These terms show up
in all equations (except in that for $\;a\;$) and influence the
motion. Integration that includes these terms gives results very
close to the Goldreich approximation (approximation (\ref{90})
that neglects the said terms and approximates the secular parts of
the nonosculating elements with those of their osculating
counterparts). However, this agreement takes place only at time
scales of order millions to dozens of millions of years. At larger
time scales, differences begin to accumulate (Lainey et al 2005).

In real life, the equinoctial-precession rate of the planet,
$\;\mubold\;$, is not constant. Since the equinoctial precession
is caused by the solar torque acting on the oblate planet, this
precession is regulated by the relative location and orientation
of the Sun and the planetary equator. This is why $\;\mubold\;$ of
a planet depends upon this planet's orbit precession caused by the
pull from the other planets. This dependence is described by a
simple model developed by Colombo (1966).


 \section{Conclusions: how we benefit from the gauge freedom.}

 In this article we gave a review of the gauge concept in orbital and attitude dynamics.
 Essentially, this is the freedom of choosing nonosculating orbital (or rotational) elements,
 i.e., the freedom of making them deviate from osculation in a known, prescribed,
 manner.\\

 {\textbf{The advantage of elements introduced in a nontrivial gauge is that in certain
 situations the choice of such elements considerably simplifies the mathematical description of
 orbital and attitude problems.}} One example of such simplification is the Goldreich (1965)
 approximation (\ref{90}) for satellite orbiting a precessing oblate planet. Although performed
 in terms of non-osculating elements, Goldreich's calculation has the advantage of mathematical
 simplicity. Most importantly, later studies (Efroimsky 2005a,b) have confirmed that Goldreich's
 results, obtained for nonosculating elements, serves as a very good approximation for
 the osculating elements. To be more exact, the secular parts of these nonosculating elements
 coincide, in the first order over the precession-caused perturbation, with those of their
 osculating counterparts, the difference accumulating only at very long time scales -- see the
 end of section 3 above. A comprehensive investigation into this
 topic, with the relevant numerics, will be presented in Lainey et
 al (2005).

 On the other hand, {\textbf{neglect of the gauge freedom may sometimes produce
 camouflaged pitfalls caused by the fact that nonosculating elements lack evident physical
 meaning}}. For example, the nonosculating ``inclination" does not coincide with the real,
 physical inclination of the orbit. This happens because nonosculating elements parameterise
 instantaneous conics nontangent to the orbit. Similarly, nonosculating Andoyer elements
 $\;L\,,\;G\,,\;H\;$ are no longer the same projections of the angular momentum as their
 osculating counterparts.
 ~\\
 ~\\

 \noindent
 {\underline{\bf{\Large{\textbf{Appendix 1.~Mathematical formalities:}}}}}
  ~\\
  ~\\
  \noindent
 {\underline{{\Large{\textbf{Orbital dynamics in the normal form of Cauchy}}}}}
 ~\\
  ~\\
Let us cast the perturbed equation
 \begin{equation}
 {\bf \ddot{
 r}}\;=\;{\Fbold}\;+\;{\Delta \efbold}\;=\;-\;\frac{\mu}{r^2}\;
 \frac{{\erbold
 }}{r}\;+\;{\Delta \efbold}\;\;\;
 \label{A1}
 \end{equation}
 into the normal form of Cauchy:
 \begin{eqnarray}
 \doterbold\;=\;{\bf{
 {v}}}
 \;\;\;\;,\;\;\;\;\;\;\;\;\;\;\;\;\;\;\;\;\;\;\;\;\;\;\;\;\;\;\;\;\;\;\;\;
 \;\;\;\;\;\;\;\;\;\;\;\;\;\;\;\;\;\;\;\;\;\;\;\;\;\;\;\;\;\;\;\;\;\;\;\;
 \;\;\;\;\;\;\;\;\;\;\;\;\;\;\;\;\,\;\;
 \label{A2}\\
 {{\bf {\dot {
 {v}}}}}\;=\;-\;\frac{\mu}{r^2}\;
 \frac{{{ \erbold}}}{r}\;+\;{\Delta \efbold}\left(\;{\erbold}(t\,,\;
 C_1\,,\,...\,,\,C_6)\;,\;\;
 {\bf{
 {v}}}(t\,,\;C_1\,,\,...\,,\,C_6)\;,t\;\right)\;\;\;.\;~
 \label{A3}
 \end{eqnarray}
 Insertion of our ansatz
   \ba
 \erbold\;=\;\efbold\left(\,t\,,\;C_1(t)\,,\;.\;.\;.\;,\;C_6(t)\,\right)\;\;\;,\;\;\;
 \label{A4}
 \ea
 will make (\ref{A2}) equivalent to
 \begin{equation}
 {\bf{
 v}} \;=\;
 \frac{\partial \efbold}{\partial t}\;+\;
 \sum_i \;\frac{\partial \efbold}{\partial C_i}\;\;
 {\dot{C}}_i\;\;\;.
 \label{A5}
 \end{equation}
 The function $\;\efbold\;$ is, by
 definition, the generic solution to the unperturbed equation
 \begin{equation}
 {\bf \ddot{
 r}}\;=\;{\Fbold}\;=\;-\;\frac{\mu}{r^2}\;\frac{{\erbold}}{r}\;\;\;.\;
 \label{A6}
 \end{equation}
This circumstance, along with (\ref{A5}), will
 transform (\ref{A2}) into
 \begin{eqnarray}
  \sum_i \;\frac{\partial
 {\bf
 g}}{\partial C_i}\;\;{\dot{C}}_i\;+\;{\bf{\dot{\Phibold}}}\;=\;
 {\Delta \bf{
 F}}\left(\;\;{\efbold}(t\,,\;C_1\,,\,...\,,\,C_6)\;\;,\;\;\;\;
 {\bf{
 {g}}}(t\,,\;C_1\,,\,...\,,\,C_6)\;+\;\Phibold\;\; \right) \;\;\;\;\;\;
 \label{A7}
 \end{eqnarray}
where  \ba
 \Phibold\;\equiv\;\sum_{i}\,\frac{\partial \efbold}{\partial
 C_j}\;\dot{C}_j\;\;\;
 \label{A8}
 \ea
 is an identity,  $\,{\efbold}(t\,,\,C_1\,,\,...\,,\,C_6)\,$ and $\,{\bf{
{g}}}(t\,,\,C_1\,,\,...\,,\,C_6)\equiv{\partial
\efbold}/{\partial t}\,$ being known functions. Now (\ref{A7} -
\ref{A8}) make an incomplete system of six first-order equations
for nine variables $\,(C_1\,,\;...\;,\,C_6\,,$
$\Phi_1\,,\;...\;,\;\Phi_3)\,$. So one has to impose three
arbitrary conditions on $\,C\,,\;\Phi\,$, for example as
 \begin{equation}
 \Phibold\;=\; {\Phibold}(\;t\,,\;C_1\,,\,...\,,\,C_6\;)\;\;\;.\;
 \label{A9}
 \end{equation}
 This will result in a closed system of six equations for six variables $\;C_j\;$:
 \begin{eqnarray}
  \sum_i \;\frac{\partial
 {\bf
 g}}{\partial C_i}\;\;{\dot{C}}_i\;=\; {\Delta \bf{
 F}}\left(\;\;{\efbold}(t\,,\;C_1\,,\,...\,,\,C_6)\;\;,\;\;\;\;
 {\bf{
 {g}}}(t\,,\;C_1\,,\,...\,,\,C_6)\;+\;\Phibold\;\; \right)
 \;-\;{\bf{\dot{\Phibold}}}
 \;\;\;\;\;\;
 \label{A10}
 \end{eqnarray}
 \begin{equation}
 \sum_i \;\frac{\partial {\bf \efbold}}{\partial C_i}\;\frac{d
 C_i}{d t}\;=\; {\Phibold}~~~,~~~~~~~~
 ~~~~~~~~~~~~~\,~~~~~~~~~~~~~~~~~~~~~~~~~~~~~~~~~~~~~~~~~~~~~~~~~~~
 \label{A11}
 \end{equation}
 $\Phibold\;=\;{\Phibold}(\;t\,,\;C_1\,,\,...\,,\,C_6\;)$ now
 being some fixed function (gauge).\footnote{~Generally, $\Phibold$ may depend also upon
 the variables' time derivatives of all orders:
 $\,{\Phibold}(t;\,C_{i}\,,\,{\dot{C}}_i\,,\,
 {\ddot{C}}_i\,,\,...)$.
 This will give birth to higher time
 derivatives of $\;C\;$ in subsequent developments and will require additional
 initial conditions, beyond those on $\;\erbold\;$ and
 $\;\doterbold\;$, to be fixed to close the system. So it is
 practical to accept (\ref{A9}).}
A trivial choice is
$\,{\Phibold}(\;t\,,\;C_1\,,\,...\,,\,C_6\;)\,=\,0\,$, and this is
what is normally taken by default.  This choice is only one out of
infinitely many, and often is not optimal. Under an arbitrary,
nonzero, choice of the function
$\,{\Phibold}(\;t\,,\;C_1\,,\,...\,,\,C_6\;)\,$, the system
(\ref{A10} - \ref{A11}) will have some different solution
$\;C_j(t)\;$. To get the appropriate solution for the Cartesian
components of the position and velocity, one will have to use
formulae
 \ba
 \erbold\;=\;\efbold(\;t\,,\;C_1\,,\,...\,,\,C_6\;)
 ~~~,~~~~~~~~~~~~~~~~~~~~~~~~~~~~~~~~~~~~~~~~~~~~~~~~~~~
 \label{A12}
 \ea
 \ba
 \doterbold\;\equiv\;{\bf{
 {v}}}\,=\,
 {\bf{
 {g}}}(\;t\,,\;C_1\,,\,...\,,\,C_6\;)\;+\;
 {\Phibold}(\;t\,,\;C_1\,,\,...\,,\,C_6\;)~~~,~~~~~~~~~~~~~~~
 \label{A13}
 \ea

~\\

~\\

 \pagebreak

 \noindent
{\underline{\bf{\Large{\textbf{Appendix 2.~~~}}}}}
 {\underline{{\Large{\textbf{Precession of the equator of date}}}}}
 ~\\
  ~\\
 \noindent
 {\underline{{\Large{\textbf{relative to the equator of
 epoch}}}}}\\
  ~\\
 The afore introduced vector
 $\;\mubold\;$ is the precession rate of the equator of date relative to the equator of
 epoch. Let the inertial axes $\;(\,X\,,\;Y\,,\;Z\,)\;$ and the corresponding
 unit vectors $\;(\,\mathbf{\hat{X}}\,,\;\mathbf{\hat{Y}}\,,\;
 \mathbf{\hat{Z}}\,)
 \;$ be fixed in space so that $\;X\;$ and $\;Y\;$ belong to the equator of
 epoch.
 A rotation within the equator-of-epoch plane by longitude $\;h_p\;$, from
 axis $\;X\;$, will define the line of nodes, $\;x\;$. A rotation about
 this line by an inclination angle $\;I_p\;$ will give us the planetary equator
 of date.
 The line of nodes $\;x\;$, along with axis $\;y\;$ naturally chosen within
 the
 equator-of-date plane, and with axis $\;z\;$ orthogonal to
 this plane, will constitute the precessing coordinate system, with the
 appropriate basis denoted by $\;(\,\mathbf{\hat{x}}\,,\;\mathbf{\hat{y}}\,,
 \;\mathbf{\hat{z}}\,)\;$.

 In the inertial basis $\;(\,\mathbf{\hat{X}}\,,\;\mathbf{\hat{Y}}\,,\;
 \mathbf{\hat{Z}}\,)
 \;$, the direction to the North Pole of date is given by
  \ba
  {\bf{\hat{z}}}\;=\;\left(\;\sin I_p\,\sin h_p\;\;,\;\;\;-\,\sin I_p\,\cos h_p
  \;\;,\;\,\;\cos I_p\;\right)^{^T}
  \label{z}
  \label{A14}
  \ea
 while the total angular velocity reads:
 \ba
 \omegabold^{(inertial)}_{total}\;=\;{\bf{\hat{z}}}\;\Omega_z\;+\;
 \mubold^{(inertial)}
 \;\;\;,
 \label{A15}
 \ea
 the first term denoting the rotation about the precessing axis
 $\;{\bf{\hat{z}}}\;$, and
 the second term being the precession rate of $\;{\bf{\hat{z}}}\;$
 relative to the inertial frame
 $\;(\,\mathbf{\hat{X}}\,,\;\mathbf{\hat{Y}}\,,\;
 \mathbf{\hat{Z}}\,)\;$. This precession rate is given by
 \ba
 \mubold^{(inertial)}\;=\;\left(\;
 {\dot{I}}_p\,\cos h_p\;\;\;,\;\;\;\,{\dot{I}}_p\,\sin h_p
  \;\;\;,\;\;\,\;{\dot{h}}_p\;\right)^{^T}\;\;\;,
 \label{muinertial}
 \label{A16}
 \ea
 because this expression satisfies $\;\mubold^{(inertial)}\;\times\;
 {\bf{\hat{z}}} \;=\;{\bf{\dot{\hat{z}}}}\;$.

 In a frame co-precessing with the equator of date, the precession
 rate will be represented by vector
 \ba
 \mubold\;=\;{\bf{\hat{R}}}_{i\rightarrow p}\;\;\mubold^{(inertial)}\;\;\;,
 \label{A17}
 \ea
 where the matrix of rotation from the equator of epoch to that of date (i.e.,
 from the inertial frame to the precessing one) is given by

 \ba
 {\bf{\hat{R}}}_{i\rightarrow p}\;=\;\;\;\;\;\left[
 \begin{array}{ccc}
 \nonumber
 ~~~~~~~~~~~~\cos h_p\;\;\;\;\;\;\;\;\;\;\;\;\;\;\;\;\;\;\;\;\;\;\;\;\;\sin h_p
 \;\;\;\;\;\;\;\;\;\;\;\;\;0\;\;\;\;\;\\
 \nonumber\\
 \nonumber
 -\;\cos I_p\;\sin h_p\;\;\;\;\;\;\;
 \;\;\;\;\;\;\;\;\cos I_p\;\sin h_p\;\;\;\;\;\;\;\;\;\;\sin I_p\\
 \label{A18}\\
 \nonumber
 ~~\;\;\,\sin I_p\;\sin h_p\;\;\;\;\;\;\;\;\;\;-\;\sin I_p\;\sin h_p
 \;\;\;\;\;\;\;\;\;\;\cos I_p\\
 \nonumber
 \end{array}
 \right]
 \ea
 From here we get the components of the precession rate, as seen
 in the co-precessing coordinate frame $\;(x\,,\;y\,,\;z)\;$:
 \ba
 \mubold\;=\;\left(\;\mu_1\;,\;\;\mu_2\;,\;\;\mu_3\;\right)^{^T}\;=\;\left(\;
 {\dot{I}}_p\;\;,\;\;\;\,{\dot{h}}_p\,\sin I_p
  \;\;,\,\;\;\;{\dot{h}}_p\,\cos I_p\;\right)^{^T}\;\;\;.
 \label{A19}
 \ea

 \pagebreak

\end{document}